\documentclass[preprint]{aastex}
\usepackage{graphicx,epsfig}
\tighten
\begin{document}
\title{Gas Dynamics in  NGC~5248: Fueling a 
Circumnuclear Starburst Ring of Super Star Clusters}
%
\author{Shardha Jogee}
\affil{Division of Physics, Mathematics, and Astronomy, MS
105-24, California Institute of Technology, Pasadena, CA 91125. 
 Email: sj@astro.caltech.edu} 

\author{Isaac Shlosman\altaffilmark{1,2}}
\affil{Joint Institute for Laboratory Astrophysics, University of 
Colorado, Campus Box 440, Boulder, CO 80309-0440. 
Email: shlosman@jila.colorado.edu}

\altaffiltext{1}{JILA Visiting Fellow}
\altaffiltext{2}{Permanent address: Department of Physics and Astronomy,
University of Kentucky, Lexington, KY 40506-0055}     

\author{Seppo Laine}
\affil{Space Telescope Science Institute, 3700 San Martin
Drive, Baltimore, MD 21218. 
Email:~laine@stsci.edu}

\author{Peter Englmaier} 
\affil{MPI f{\" u}r extraterrestrische Physik, Postfach 1312, 85741 Garching,
Germany. Email:~ppe@mpe.mpg.de}

\author{Johan H. Knapen}
\affil{Isaac Newton Group of Telescopes, Apartado 321, Santa
Cruz de La Palma, E-38700 Spain,  and 
Department of Physical Sciences, University of 
Hertfordshire, Hatfield, Herts AL10 9AB, UK. 
Email: knapen@ing.iac.es}

\author{Nick Scoville}
\affil{Division of Physics, Mathematics, and Astronomy, MS 
105-24, California Institute of Technology, Pasadena, CA 91125. 
Email: nzs@astro.caltech.edu}

\and

\author{Christine D. Wilson} 
\affil{Department of Physics and Astronomy, McMaster University,
Hamilton Ontario L8S 4M1, Canada. Email: wilson@physics.mcmaster.ca}


\author{\space}
 
\begin{abstract} 

Through observations and modeling, we demonstrate how the recently 
discovered  large-scale bar in NGC~5248 generates spiral structure 
which extends from 10 kpc down to 100 pc, fuels star formation  on 
progressively smaller scales, and drives disk evolution. Deep inside 
the bar, two massive molecular spirals cover nearly  $180^\circ$ in 
azimuth, show streaming motions of 20--40~km~s$^{-1}$, and feed a starburst 
ring of  super star clusters at 375 pc. They also connect to two narrow 
$K$-band spirals which delineate the UV-bright star clusters in the ring. 
The data suggest that the $K$-band spirals are young, and the starburst 
has been triggered  by a  bar-driven spiral density wave (SDW). The 
latter may even have propagated closer to the center where a second 
H$\alpha$ ring and a dust spiral are found. The molecular and \it HST \rm 
data support a scenario where stellar winds and supernovae efficiently clear
out gas from dense star-forming regions on timescales less than a few Myrs.
We have investigated the properties of massive CO spirals within the framework 
of bar-driven SDWs, incorporating the effect of gas  self-gravity. We find 
good agreement between the model predictions and the observed morphology, 
kinematics, and pitch angle  of the spirals. This combination of observations 
and modeling  provides the best evidence to date for a strong dynamical coupling 
between the nuclear region and the surrounding  disk. It also confirms that a 
low central mass concentration, which may be common in late-type galaxies, 
is particularly favorable to the propagation of a bar-driven gaseous SDW deep 
into the central region of the galaxy, whereas a large central mass concentration 
favors other processes, such as the formation and decoupling of nuclear bars.
\end{abstract} 

\section {Introduction}

The level of activity and the rate of evolution in the circumnuclear region 
of galaxies depend  to a large degree on dynamical processes which transport
molecular gas. Stellar bars and other non-axisymmetries are destined to drive
the gas towards the central kpc (Simkin, Su, \& Schwarz 1980; Shlosman, 
Frank, \& Begelman 1989; Athanassoula 1992). In some cases, stellar bars can
be induced by tidal interactions (e.g., Noguchi 1988). The secular evolution
of disk galaxies, therefore, proceeds in the direction of increasing central
mass concentration in the baryonic material (e.g., Pfenniger 1996). 

The dynamical and secular evolution in the central kpc,  including stellar 
and non-stellar activities,  depend on the degree of non-axisymmetry in 
this region. Strongly non-axisymmetric structures   will increase 
the rate of evolution by facilitating the  transfer of angular momentum and 
mass (e.g., Shlosman 2001). Observations of the central regions in barred galaxies
provide clear evidence for non-axisymmetries such as nuclear spiral
structure, triaxial bulges, and nuclear (secondary) bars (e.g., Laine et al.
1999a; Jogee, Kenney, \& Smith 1999; Regan \& Mulchaey 1999; Martini \& Pogge
1999;  Bureau 2002; Laine et al. 2002). Numerical simulations do reproduce
much of this morphology, but fundamental issues remain, such as the state of
the molecular gas in the central kpc, the formation and dynamical decoupling of
nested bars, the stellar and gas dynamics in decoupled systems, and  the 
interactions of bars with dark matter halos (e.g., Friedli \& Martinet  1993; 
Jogee, Kenney, \& Smith 1999; Maciejewski \& Sparke 2000; H\"uttemeister 
\& Aalto 2001; Shlosman \& Heller 2002; Athanassoula 2002; Shlosman 2002). 

Both theory and observations reveal the importance of inner Lindblad
resonances (ILRs) in slowing down the radial gas inflow  within the 
inner kpc of barred galaxies  (e.g., Schwarz 1984; Combes \& Gerin 1985; 
Shlosman et al. 1989; Athanassoula 1992; Sofue 1991; Kenney et al. 1992; 
Knapen 1995a,b; Jogee 2001; Jogee et al. 2001). The resulting   concentration 
of  molecular gas leads to the formation of nuclear rings on scales of a few 
hundred pc (e.g., review by Buta \& Combes  1996). These rings  exhibit elevated 
rates of star formation (SF) due to the gas becoming  gravitationally unstable 
(e.g., Elmegreen 1994; Knapen et al.1995b; Jogee 1999; Jogee et al. 2001), and 
are dynamically important (Heller \& Shlosman  1996). 

The nearby ($D=15$ Mpc) grand-design  spiral galaxy NGC~5248 is a great
laboratory  for investigating gas dynamics, barred and spiral morphologies,
and their impact on circumnuclear SF. It shows spectacular
grand-design spiral structure which extends over a large dynamic range in 
radius, from $\sim 10$~kpc to 100 pc, and winds through  a large angle
($> 360^\circ$) in azimuth. NGC~5248, with a revised Hubble type SAB(rs)bc
(de Vaucouleurs et al. 1991), was until recently considered to host a short
bar with a semi-major axis of $22\arcsec$ (1.6 kpc; Martin 1995), or even
treated as unbarred (e.g., Patsis et al. 1997). However, Jogee et al. (2002,
hereafter Paper~I) have shown from a deep $R$-band image (Fig.~1), as well as
from dynamical and photometric analyses,  that the spiral structure is being
driven by  an extended, moderately strong  stellar bar which has a deprojected 
ellipticity of 0.44 and  a  semi-major axis $a_{\rm bar}$ of $\sim 
95\arcsec$ (7.1~kpc). The bar is embedded  within a fainter outer disk 
which is visible out to a radius of $230\arcsec$ (17.2~kpc). 
Previous optical and near-infrared (NIR) images did not have  the sensitivity and/or  
field of view to clearly detect the outer disk, and therefore the outer 
region of the bar was mistaken for an inclined outer disk. 
The corotation  radius ($r_{\rm CR}$) of the bar lies at $\sim 115\arcsec$ (8.6~kpc), 
assuming the empirical relationship $r_{\rm CR} = (1.2\pm 0.2) a_{\rm bar}$ 
(Athanassoula 1992). Table~1 summarizes the parameters adopted for NGC~5248 in this paper.
The inner kpc of the bar hosts a well known ring of ``hot spots''
(e.g., Buta \& Crocker 1993; Kennicutt, Keel, \& Blaha 1989; Elmegreen et al. 
1997)  which has been resolved into bright
H{\sc ii} regions and young super star clusters (SSCs)  by {\it HST} broad-band,
H$\alpha$, and Pa$\alpha$ images (Maoz et al. 2001). Inside this ring resides
a second nuclear H$\alpha$ ring with a radius of $1 \farcs 25$  (Laine et al.
2001; Maoz et al. 2001), and a nuclear dust spiral (Laine et al. 1999a)
between $1\arcsec$ (75~pc) and $4\arcsec$ (300~pc).                 

High resolution observations of  molecular gas can provide important
constraints on the dynamics and evolution of the central regions in disk
galaxies. Not only  does the molecular gas fuel the ongoing SF, 
but its morphology and kinematics can also be a powerful tracer of
non-axisymmetric features in the underlying galactic potential (e.g., Knapen
et al. 1995b; Sakamoto et al. 1995; Jogee, Kenney, \& Smith 1999; Knapen et
al. 2000). We present new high resolution ($1\farcs 9\times 1 \farcs 4$ or
$150 \times 100$~pc) Owens Valley Radio Observatory (OVRO) CO ($J$=1--0)
observations of NGC 5248, and compare the data with Fabry-Perot H$\alpha$ 
observations (Laine et al. 2001), optical and NIR broad-band images, archival
{\it HST} images (Maoz et al. 2001), and theoretical models.
The main sections in this paper are organized as follows. $\S$ 2 outlines 
the observations and data reduction. $\S$ 3 describes the stellar, dust,
and  gaseous grand-design spirals in NGC~5248. In $\S$ 4, \rm we  present the
CO and $K$-band data, showing the striking molecular and young stellar spirals.
We describe  the CO kinematics and present the rotation curve of NGC~5248 in
$\S$ 5. In $\S$ 6, we discuss evolution in the inner kpc of NGC~5248,
addressing in particular the feeding of the starburst ring with cold
molecular fuel, the local SF  properties based on different
tracers,  and the interplay  of the SSCs with the interstellar medium.
In $\S$ 7.1, estimates of the number and locations of the ILRs in NGC~5248
are presented. To account for the observed properties, the data are
compared to new revised theoretical models in $\S$ 7.2. In $\S$ 8, we discuss the 
implications of this work for ISM redistribution and galaxy evolution. 

\section {Observations} 

\subsection {CO interferometric observations}

The central $65\arcsec$ of NGC~5248 were observed in the  redshifted CO
($J$=1--0) transition at  114.83~GHz  
with the OVRO millimeter-wave array  (Padin et al. 1991) between
March 1999 and Feb. 2001. The array consists of six 10.4~m telescopes with
a primary half power  beam width of $65\arcsec$ at 115~GHz. The galaxy was
observed  in five array configurations  which range from compact to
ultra-wide,  and  include projected baselines  from 12 to 483~m.  
Data were obtained simultaneously with an analog continuum correlator of
bandwidth 1~GHz and  a digital spectrometer set-up  which produced four
independent modules  that each have 32 channels with a velocity resolution of
5.2~km~s$^{-1}$.  For our observations, the modules were partially overlapping
and covered a  total bandwidth of 240 MHz (600~km~s$^{-1}$) with 116 frequency
channels.  We corrected for temporal phase variations by interspersing 
integrations on the galaxy with observations of  a phase calibrator,
typically a quasar, every 20--25 minutes.  Passbands were calibrated on the
bright quasars  3C273, 3C84, and 3C345. The absolute flux scale was determined 
from observations of  Uranus, Neptune, and 3C273, and has an 
accuracy of  20\%. The passband and flux calibration of the data were
carried out using the  Owens Valley millimeter array software  (Scoville et
al. 1993).  We used the NRAO AIPS software to map  the calibrated
$uv$ data and to deconvolve the channel maps with the CLEAN algorithm
and robust weighting as implemented in the AIPS task  IMAGR. 
Primary beam correction was applied. The parameters of the channel maps   are
summarized in Table~2. 
Emission above the $3\sigma$ level is detected in 55
channels  in the velocity  range of 1012 to   1293~km~s$^{-1}$. 
We combined the cleaned channel maps showing emission to make moment 0, 1, 
and 2 maps  which represent the total intensity, the intensity-weighted
velocity field,  and the  velocity dispersion field, respectively. Maps with
different Briggs robustness parameters ranging from  -7 to 4 were made to
allow for different weighting schemes that  compromise between resolution and
signal-to-noise. The  maps discussed here 
contain a  total flux of 385 $\pm$ 20 Jy km~s$^{-1}$ over the central 
$38\arcsec$,  corresponding to  85\% of the single dish 
flux in the central $45\arcsec$ (Young et al. 1995).

\subsection{Near-infrared and optical data}

We complement the CO observations with  optical and NIR 
data which are summarized in Table~3. These include   published 
archival  {\it HST}  broad-band  and H$\alpha$+[N{\sc ii}] images,   
TAURUS Fabry-Perot H$\alpha$ data (Laine et al. 2001), and $J-K$ color maps. 
We also obtained  images of NGC~5248 through  the $R$, $B$,  and $K_{\rm s}$ filters.

NGC~5248 was imaged through  the Harris $R$-band  filter for a total exposure
time of 30 minutes using the Wide Field Camera on the 2.5~m Isaac Newton
Telescope (INT) in La Palma  in August 2001. The data frames had a plate scale
of $0\farcs 33$/pix, a  field of view of $11\farcm 3 \times  22\farcm5$, and an
average seeing of  $1\farcs8$. Bias-subtracted and  flat-fielded frames were
obtained from  the INT  data reduction pipeline. Sky-fringing at the  1\%
level is present, an effect known for the  thinned EEV chip CCD4.  We used the
IRAF package for 
fixing bad pixels, cleaning cosmic rays, and convolving the frames 
to a common seeing before combining them into a final image. After masking
out stars and faint background galaxies in the outer parts of the final
image,  isophotal analysis of the $R$-band light  was performed with the 
``isophote'' package in IRAF. Details are presented in Paper~I.

A  $B$-band  image of NGC~5248, taken with  the Prime Focus Camera on 
the 2.5~m Isaac Newton Telescope with a total exposure time of 20 minutes, 
was retrieved from the Isaac Newton Group archive. The image has 
$0\farcs 589$ pixels  and a field of view of $11\arcmin$.
We used the IRAF package in the standard way for data reduction.
NGC~5248 was also imaged through  the $K$-short ($K_{\rm s}$) filter  for  a total
on-source exposure time of 12 minutes using  the INGRID camera on the 4.2~m
William Herschel Telescope. 
Seeing was $0 \farcs 8$. On-source  frames were
interspersed with off-source frames  pointed on a nearby blank area of the sky
for reliable sky subtraction. The final reduced images have $0\farcs 24$ pixels 
and a $5\farcm 1$ field of view. 
The data acquisition and reduction are described in more detail in Knapen et
al. (in preparation). 

\section {The grand-design spirals in NGC~5248}

NGC~5248 exhibits spectacular  grand-design spiral structure 
in all disk components, namely  gas, dust, and young stars.
Spiral arms exist from  $\sim$~10~kpc to  $\sim$~100~pc 
and cover more than $360^\circ$ in azimuth, as illustrated by 
Figs.~1-7. 
The  $R$-band  (Fig.~1)  and $B$-band  (Fig.~2) images  show 
two bright, relatively symmetric   
spirals which are particularly prominent on the leading
edge of the large-scale stellar bar between $30\arcsec$ and 
$90\arcsec$.   These stellar spiral arms exhibit dust lanes on their inner
(concave) side out to at least $70\arcsec$ (Fig.~2 and  Paper~I), 
as expected inside the corotation resonance (CR). 
The offset between the dust and young stars is consistent 
with the view that shocks, seen as dust lanes along the leading edges of a  
moderately strong  bar,  compress gas to form massive young stars. 
SF in the bar of NGC~5248 is exceptionally strong, even in the NIR (Paper~I). 
Patchy $K$-band spirals (Fig.~3) accompany the $B$-band spirals on 
the leading edge of the bar, and peaks  in  $K$-band emission 
coincide with H{\sc ii}  regions  in  the large-scale H$\alpha$ image.  
The strong SF along the moderately strong  bar in NGC~5248 
is consistent with the  current understanding of gas flows and  SF  in bars.  
Namely, in strong bars which exhibit almost straight  
offset  dust lanes believed to trace strong shocks, the  strong shear in 
the postshock flow can inhibit SF  (e.g., Athanassoula 1992).  
However, in weaker bars with curved offset dust lanes, 
the weaker shocks and shear can induce SF  rather than 
inhibit it. For instance,  the collapse of gas cores to form stars can be 
induced by weak shocks with speeds of order 20--30 
km~s$^{-1}$  (e.g., Vanhala \& Cameron 1998). 
M100 (e.g., Elmegreen et al. 1989; Knapen \& Beckman 1996), 
NGC~2903 (e.g., Sheth 2001),  NGC~4254, and NGC~4303  (e.g., Koopmann 1997)  
are but a few examples of systems with significant SF  along a weak bar.  
While in many  strongly barred galaxies  optical spiral 
arms are prominent outside the bar but not within it, 
NGC 5248 illustrates how  intense SF  along a 
fairly weak  bar can lead to conspicuous open spiral arms 
within the bar itself.

The $R$-band spirals cross the bar major-axis around 
$95\arcsec$  where the bar ends.  This radius likely denotes the location 
of the  ultraharmonic resonance where  large-scale bars 
tend to end  due to the onset of chaos and orbital instabilities. 
Secondary faint spirals stem off the bar and become particularly
prominent from  $115\arcsec$ to $150\arcsec$. These may be related 
to secondary compression zones  which arise near CR (e.g., Buta 1984).  
Between  $150\arcsec$ and $230\arcsec$, very faint spirals  
extend away from the bar end and cover $90^\circ$--$100^\circ$ in azimuth 
(Fig. 1) in the faint outer disk. 
The outer Lindblad resonance (OLR), where spiral arms are expected to end,
is probably around $230 \arcsec$.

The spiral structure in NGC~5248 not only extends out to large radii,
but also continues towards the central regions of  the bar in 
three components, namely gas, dust, and young stars. The $R$-band (Fig.~1), 
$B$-band (Fig.~2),  $K$-band (Fig.~3), and $I-K$ (Fig.~4a) spirals can be 
readily followed from $70\arcsec$ to $26\arcsec$  where they cross the 
bar major axis. Interior to this point, the continuity in spiral structure 
is especially noticeable in the dust spirals revealed by the  $I-K$  (Fig.~4a)  
and   $J-K$ (Fig.~4b; see also P\'erez-Ram\'\i rez et al. 2000) color maps.
The dust spirals  appear  continuous  from  $70\arcsec$ to  $5\arcsec$, 
and  cover nearly another $180^\circ$ in azimuth between $26\arcsec$ 
and  $5\arcsec$ before they join a circumnuclear starburst ring. The 
CO (1--0)  maps described in $\S$ 4 show that the dust spirals are associated 
with  two massive molecular spiral arms which cover at least $180^\circ$ 
in azimuth between  $20\arcsec$ and  $5\arcsec$ (Fig. 5) and then connect to 
the starburst ring of  H{\sc ii} regions (Fig.~6) and UV-bright SSCs 
(Fig. 7). It is not entirely clear if the $K$-band spirals 
extend continuously from  $70\arcsec$ to  $5\arcsec$ as do the dust 
spirals: faint  patches of $K$-band emission can be seen along the 
western molecular spiral between $19\arcsec$ and $12\arcsec$, and  at  
$\sim$~$8\arcsec$  two striking  bright inner $K$-band spirals (Fig. 8) 
connect to the molecular spirals. The narrow southern $K$-band spiral is 
particularly conspicuous. The $K$-band spirals continue from 600 pc to 
$\sim$~$3\arcsec$ (225 pc), and delineate the UV-bright SSCs as they 
cross the starburst ring (Figs. 7 and 8). Interior to the starburst ring  
an adaptive optics  $J-K$ color map  reveals a grand-design nuclear dust 
spiral  (Fig.~9) which can be traced from $\sim$~$4\arcsec$ (300~pc) 
down to about $\sim$~$1\arcsec$ (75~pc; Laine et al. 1999a). This dust spiral 
appears to bridge  the $K$-band spirals and the second nuclear H$\alpha$ 
ring of  radius 95~pc where double peaks of CO emission can be seen ($\S$ 4). 
To account for the observed morphology, the data will be  compared to 
theoretical models in  $\S$ 7, after a closer look at the molecular gas and 
SF properties in $\S$ 4--6.

\section {The circumnuclear CO and  $K$-band morphology} 

The  OVRO CO (1--0) total intensity map  (Fig.~4) has a significantly 
better resolution ($1 \farcs 9 \times 1 \farcs 4$  or $140 \times 100$~pc) 
than the previously  published  ($4\farcs 4 \times 3\farcs 7$ or $330 
\times 280$~pc) Nobeyama Millimeter Array (NMA) map (Sakamoto et al. 1999). 
This high resolution map emphasizes the strongest features and may 
miss some of the fainter diffuse emission. Several striking features 
can be identified.  Two  relatively bisymmetric  trailing  CO spiral arms 
can be followed from a radius of  $20\arcsec$ (1.5~kpc) to $5\arcsec$ (375~pc). 
The arms, also visible in the NMA map, are  now resolved into multiple  clumps 
with sizes of  $2\arcsec$ to $4\arcsec$  (150--300 pc) and masses in the range 
of  several times 10$^6$--10$^7$ M$_{\sun}$, comparable to giant molecular
associations (e.g, Vogel, Kulkarni, \& Scoville 1988). The CO arms  connect 
to the circumnuclear starburst ring  of H{\sc ii} regions (Fig.~6) and SSCs 
(Fig.~7) at a radius  of $5\arcsec$.
Inside  this ring resides a double-peaked CO feature labeled F1 in Fig.~5. 
Its two CO peaks lie partially  on the inner nuclear H$\alpha$ ring of radius 
$1 \farcs 25 $ (95~pc) and are crossed by the arms of the grand-design nuclear 
dust spiral (Fig.~9; Laine et al. 1999a).  Fainter CO emission  is present in 
the starburst ring, northwest and southeast of the CO peaks. It is unclear 
if the apparent bridges between the faint CO emission and the two massive CO 
arms or the CO feature F1 are real or simply the result of  beam-smearing.

The massive CO arms are associated with prominent dust spirals (Fig. 4b) 
delineating shocks. These dust spirals form part of a continuous 
grand-design  spiral which can be followed from  $70\arcsec$ to at least 
$5\arcsec$  and  crosses the  bar major-axis at $\sim$~$26\arcsec$. 
The $K_{\rm s}$ image  of $0 \farcs 8 $  resolution  (Fig.~8) shows some 
faint patches of emission  along the western CO arm between 
$12\arcsec$ and $19\arcsec$  and  two bright
inner spirals  which connect to the  CO arms around  $\sim$~$8\arcsec$. 
The $K$-band arms cross the starburst ring while delineating the 
UV-bright SSCs (Fig.~8).  The association between the southern $K$-band 
arm and the SSCs labeled SC11, SC12,  SC13,  and SC14  in Fig.~7 is 
particularly striking. The   emission peaks  in the $K$-band  arms are likely 
to be dominated by very young supergiants, 8--10 Myr old,  which are often 
present in young star-forming regions (e.g., Leitherer \& Heckman 1995).  
The presence of such a young stellar population in  the $K$-band arms
is  consistent with the study of Maoz et al. (2001) who estimate, 
from spectral synthesis,  an age of 10--40 Myr for most of the bright SSCs. 
A young age for the $K$-band arms is also suggested by the narrow 
width of $0 \farcs 6 $ (110 pc)  of the eastern $K$-band arm. 
For an assumed stellar velocity dispersion of 10  km~s$^{-1}$, 
this width  suggests  an age of order 10--20 Myr for the arms. 
Taken together, the morphology of the CO arms, dust spirals, 
young $K$-band arms, and SSCs suggests that SF  in the starburst ring 
has been triggered by a bar-driven density wave.

The $K$-band arms are located  just outside a  very weak oval feature of 
radius $\sim$~$3\arcsec$ (Fig. 8).  This feature has a deprojected  
ellipticity of 0.1--0.2 at a position angle (PA) of $110^\circ$--$120^\circ$, 
slightly offset from the estimated PA of the line of nodes  $105^\circ$ 
(Fig. 10a and  Paper I).  This feature could be a disk-like component, 
a  late-type  bulge, or an unresolved nuclear bar. Interestingly, 
the grand-design  nuclear  dust  spiral (Fig.~9) which continues from 
$3\arcsec$ to $1\arcsec$ appears to bridge the end of the  $K$-band arms 
around $3\arcsec$ and the nuclear  H$\alpha$ ring of  radius 
$1 \farcs 25 $. The dust spiral also crosses the two CO peaks of feature F1.
At the current resolution, it is not clear whether the CO peaks of F1 
are part of  a molecular spiral associated with the dust spiral or 
whether  they are associated with a molecular ring coincident with the 
inner H$\alpha$ ring.

We  computed  the mass of molecular hydrogen in our CO (1--0) maps using 
the relation (Kenney \& Young 1989; Scoville \& Sanders  1987):   

\begin{equation} 
\frac {M_{\rm H2}} {M_{\tiny\sun}} 
= 1.1 \times 10^4 \ \left(\frac {\it X}{2.8 \times 10^{20} }\right)\ D^2 \
\int{S_{\tiny \rm CO}}\rm{d} V 
\end{equation}
where D is the  distance  in Mpc,   $ \int{S_{\tiny \rm CO}}dV$ 
is the integrated line flux  in Jy km~s$^{-1}$ and 
$X$  is the  CO--H$_{\rm 2}$ conversion factor. 
The value of $X$   in the inner Galaxy,  namely 
$2.8 \times 10^{20}$  H$_{\rm 2}$ cm$^{-2}$ K (km~s$^{-1}$)$^{-1}$ 
(Bloemen et al. 1986),  is generally referred to as the 
``standard'' value. 
For  an assumed  standard value of the conversion factor, the two 
CO spiral arms host a total of 
$9.0 \times 10^8$ M$_{\sun}$ of molecular  hydrogen,  while the CO 
feature F1 hosts $1.0 \times 10^8$ M$_{\sun}$. Including the 
contribution of helium and heavier elements for a solar metallicity (Z=0.02, 
X=0.72, Y=0.26), gives a   molecular gas content of $1.2 \times 
10^9$ M$_{\sun}$ and $1.4 \times 10^8$ M$_{\sun}$, respectively.
We estimate the enclosed dynamical mass using the circular rotation curve 
described in $\S$ 5. In the central $20\arcsec$ (1.5 kpc) radius  
molecular gas makes up 
11\% of the enclosed  dynamical mass ($1.2 \times 10^{10}$ M$_{\sun}$). Within
the central $5\arcsec$ (375~pc) radius, the CO feature F1 makes up 7\% of the
dynamical mass ($2.1 \times 10^9$ M$_{\sun}$).
Our mass estimates suffer from the usual caveats of assuming a 
standard value for the conversion factor.  
One may justifiably  question this assumption 
since the conversion factor may depend on many parameters 
such as  the dust column density, the ambient radiation field, 
the physical conditions  in the gas,  the optical thickness of the line, 
and the metallicity  of the gas.
These parameters could potentially show  significant variations between  
different regions of galaxies. 
Furthermore, multiple-line studies and radiative transfer models 
(Wall \& Jaffe 1990; Wild et al. 1992; Helfer \& Blitz 1993; Aalto et al. 1995) 
have suggested lower values of $X$   in the centers of some starburst  galaxies. 
However, several mitigating factors may help reduce deviations of 
the conversion factor from the standard value  in the central region 
of NGC~5248. While circumnuclear  star-forming regions very likely have 
higher temperatures and densities than Milky Way clouds, the effects of elevated
temperatures and densities may partially cancel  each other since $X$ depends
on ($\rho^{0.5}$/$T_{\rm B}$) for clouds  in virial equilibrium  
(e.g., Scoville \& Sanders 1987; Solomon et al. 1987), where  
$T_{\rm B}$ is the CO brightness temperature averaged over the cloud 
and $\rho$ is the density. Furthermore, most of the CO emission in NGC~5248 
is not associated with the  ring of H{\sc ii} regions and SSCs ($\S$ 6) 
where the radiation field from young  stars might heat up the gas significantly.  
As far as metallicity is concerned, the circumnuclear region of an Sbc spiral 
such as  NGC~5248 is likely to have a metallicity of at least solar,  like 
most massive spirals (e.g., Vila-Costas \& Edmunds 1992). Over this regime of
metallicity, the CO line is likely optically thick, and we expect at most a
weak dependence of the CO--H$_{2}$  conversion factor on metallicity 
(e.g., Elmegreen 1989).

%
%

\section {The molecular gas kinematics}

Fig.~11a shows the intensity-weighted CO velocity field of NGC~5248. The
spectral resolution is 5.2 km~s$^{-1}$.   
The position angle of the line of nodes ($105^\circ$) 
determined  from the outer disk in the $R$-band image (Paper~I) is marked. 
The isovelocity contours deviate from a typical ``spider diagram'' which
characterizes purely circular velocity fields. Non-circular streaming motions
are especially visible  in  the CO arms where the isovelocity contours show
significant curvatures.   
Non-circular motions are also apparent in the faint CO emission  
which lies around the two CO peaks of feature F1, near the ring 
of SF. In fact, the regions of non-circular motions 
are intersected  by the inner $K$-band spiral arms as they 
extend from  $3\arcsec$ to $8\arcsec$, crossing  the starburst ring.

At any general point, the line of sight velocity is a superposition  
of  circular motions ($\Theta_{\rm c}$), azimuthal streaming motions
($\Theta_{\rm s}$), and  radial streaming motions  ($\Pi$) in the plane, as
well as velocities ($Z$) out of the plane. 
It is particularly instructive to look at position--velocity (p--v) 
cuts along the kinematic minor and major axes of the galaxy,  
where the components 
$\Theta_{\rm s}$ and $\Pi$, respectively,  tend to zero. 
The p--v  cut along the optical minor ($15^\circ$) axis is shown in Fig.~11b. 
At a distance of $6\arcsec$ to $8\arcsec$ along the minor 
axis, where the p--v cut crosses the CO arms, the velocities to the
NE and SW show, respectively,  blueshifts and redshifts of 15--25 km~s$^{-1}$  
with respect to the systemic velocity (1153 km~s$^{-1}$). 
These velocities correspond to  local  radial streaming motions  
($\Pi$) of (15--25/sin~$i$) km~s$^{-1}$ or 25--40  km~s$^{-1}$,  if we assume 
the optical minor axis  is close to the kinematic minor axis, and the 
out-of-plane $Z$ component of velocity is  negligible. 
The radial streaming motions  correspond to inflow 
if we assume the NE side of the galactic disk is the far 
side -- an orientation  justified  at the end of this section.
We caution, however, that local radial inflow motions do not directly 
translate into a net inflow rate (see $\S$ 6).

The p--v cut (Fig.~11c) along the galaxy major axis or line of nodes 
($105^\circ$) first crosses the innermost part of the double peaked 
CO feature F1, and  then  intersects  the two CO arms.  In the inner 
regions  where we expect the rotation  curve to be steeply rising,  
the p--v  diagram  shows a turnover velocity of (115/sin~$i$) km~s$^{-1}$ 
at a $7\arcsec$ radius. 
This corresponds to 179  km~s$^{-1}$  for a revised inclination 
of $40^\circ$ adopted from Paper~I. 
Between  $8\arcsec$ and  $11\arcsec$ along  the major axis, 
there is a dip in the mean CO  velocity and an 
increase in the CO linewidth. This happens when  the p--v cut first 
intersects the CO arms  (Fig.~11a). The curvatures in the 
isovelocity contours at that location  suggest that the kinematic 
behavior in Fig.~11c  results from  non-circular azimuthal 
streaming motions. 
Further out,  from  $12\arcsec$ to $18\arcsec$ along the major axis 
the p--v cut crosses  the two CO arms, and the line-of-sight velocity 
rises by about 40 km~s$^{-1}$. 
Again, the strong curvatures in the isovelocity contours (Fig.~11b) 
suggest this rise is in large part due to non-circular 
streaming motions.

Fabry-Perot H$\alpha$ observations of NGC~5248 with a  spectral resolution 
of 20.0 km~s$^{-1}$ have been presented by  Laine et al. (2001). 
They derive a best-fit model  ``circular'' velocity in the central 
$10\arcsec$ 
by  minimizing the residual velocity difference between the observed  
Fabry-Perot H$\alpha$ velocity field and the model velocities. 
In the presence of strong non-circular motions, the best-fit circular 
velocity derived using  this method is  still contaminated by 
non-circular components. 
Their  best-fit  circular rotation curve derived assuming  a line of nodes 
of $105^\circ$ and an inclination of  $40^\circ$ is  shown on Fig.~12. 
The Fabry-Perot H$\alpha$ circular 
velocity of  (117/sin~$i$) km~s$^{-1}$ at a $5\arcsec$ radius 
is consistent with the CO velocity of  (115/sin~$i$) km~s$^{-1}$ at that 
location.  
Around  $8\arcsec$--$10\arcsec$, the  H$\alpha$ velocity field shows residual
non-circular motions of 30 km~s$^{-1}$ to  the northeast, and 15 to 20  
km~s$^{-1}$ to the south and west. These non-circular motions 
occur near the intersection point of the $K$-band arms and CO arms
with the starburst ring.

CO (1--0) observations with a lower resolution  ($5\arcsec$--$6\arcsec$) but a 
larger field of view  ($2\arcmin$) have been carried out with the 
Berkeley Illinois Maryland Array (BIMA), as part of the  
Survey of Nearby Galaxies (SONG; Regan et al. 2001). 
The data  (Das et al. in prep.)  show molecular gas 
along the large-scale bar out to  $90\arcsec$, as well as in the 
circumnuclear regions.  A best-fit ``circular'' rotation curve derived 
from the BIMA data was kindly provided to us by M. Das prior to 
publication.  It  was derived with a technique of 
minimizing residuals, similar to the method applied by Laine et al. (2001)  
to the Fabry-Perot H$\alpha$ data.  It suffers, therefore, 
from the same caveats of possible contamination by non-circular motions.
We used the high resolution  
Fabry-Perot H$\alpha$ and OVRO CO (1--0)  ``circular'' rotation curve 
from 0 to $9\arcsec$, and the BIMA data of lower resolution from 
$13\arcsec$ to $80\arcsec$ to derive an overall rotation curve  
from the center to $80\arcsec$.  This is shown in Fig. 12  where 
values are already corrected for an assumed inclination of  $40^\circ$.
We did a  cubic  spline fit  to interpolate between data points.
The turnover velocity of  (117/sin~$i$) or 182 km~s$^{-1}$ at a 
$7\arcsec$ radius is visible. 
Based on the above discussions  of Fig. 11c, we suspect that 
the dip in velocity from $8\arcsec$ to $11\arcsec$, as well as the 
40   km~s$^{-1}$ rise in velocity from $10\arcsec$ to $20\arcsec$, 
are largely  caused by non-circular motions.  

Finally, we justify the assumption  made earlier that the 
NE side of the disk is the far side. The stars in the disk must be rotating 
in a clockwise sense  if the large-scale stellar spiral arms are trailing, 
as is commonly the case with large-scale spirals. We can also
infer a clockwise sense of rotation for the stars  from the  following line of
arguments. If we start on the bar major axis and move in a clockwise sense, we
first cross  the  dust lanes on the leading  edge of the bar, then the
quasi-coincident H$\alpha$, $K$-band,  and $B$-band spirals (Paper~I). 
This suggests that  shocks, which we see as dust lanes  along the leading edges 
of the bar, compress the gas to form  massive young stars which move clockwise 
faster than the pattern speed,  and lead to the spirals made of young stars 
a little further away.  
Such offsets between dust, gas, and stars have also been seen in observations of 
spiral arms  in disks (e.g., Allen, Atherton, \& Tilanus 1986; 
Tilanus \& Allen 1989; Tilanus \& Allen 1991). 
If the gas in the central region is rotating clockwise like the stars, then
the blueshifts and redshifts on the NW and SE side of the major axis, 
respectively (Fig.~11a) imply that the NE side of the disk is the far side.
This orientation is consistent with the dust lanes in the $J-K$ map (Fig.~4b)
being redder on the western side than on the eastern side.

\section {Evolution in the inner kpc of NGC~5248 }



We now take a closer look at  evolution in the inner kpc of NGC~5248 
using the information synthesized from our high resolution 
CO (1--0), $J-K$, and $K_{\rm s}$  observations, along with published 
$HST$ images  (Maoz et al. 2001). 
In particular, we focus here on the question 
of feeding the  starburst ring  with cold molecular fuel, 
the local SF properties,   and the interplay  of the 
SSCs with the interstellar medium.

The integrated  H$\alpha$+[N{\sc ii}] luminosity  in the inner kpc 
radius of NGC~5248, after an extinction correction based 
on a dust screen model, is $\sim$ $2 \times 10^{41}$ ergs  s$^{-1}$ 
(Maoz et al. 2001). 
It  corresponds to  a current SFR of 2.1 $M_{\sun}$ yr$^{-1}$,  
assuming an extended Miller-Scalo IMF,  stellar masses from  0.1  
to 100  $M_{\sun}$, and a case B recombination.  
The true SFR could be higher if there is heavy internal 
extinction within  the H{\sc ii} regions, but its upper 
limit is constrained by the  global SFR of 3.2  $M_{\sun}$  
yr$^{-1}$, estimated for the entire galaxy. The latter rate is derived  
from  the total far infrared  luminosity  and low resolution radio 
continuum data, following the prescription of Condon (1992). 
Table 4 summarizes SFR estimates based on  different tracers. 
Our estimate of  2.1 $M_{\sun}$ yr$^{-1}$ is also consistent with 
other circumnuclear properties.
An average massive SFR  of order   0.5 $M_{\sun}$ yr$^{-1}$ or a 
total average SFR of 1.5  $M_{\sun}$ yr$^{-1}$ is suggested by the 
young age (10--40 Myr; Maoz et al. 2001) and  total mass 
(several times 10$^6$  $M_{\sun}$; Maoz et al. 2001) of the 
optically-visible stellar  clusters. The cluster ages are consistent 
with the age  (8--10 Myr)  of the $K$-band arms delineating the brightest
SSCs ($\S$ 4).  A similar  average  circumnuclear SFR of 
1.5 $M_{\sun}$ yr$^{-1}$ 
was derived by Elmegreen et al. (1997) based on the ages   
and masses 
of  ``hot spots''  estimated from  lower resolution NIR data.  
%
%

The $HST$ H$\alpha$  (Fig. 6)  image shows that most of the H$\alpha$ 
emission  originates from the circumnuclear ring of H{\sc ii} regions 
at a radius of $5\arcsec$ and the nuclear  H$\alpha$  ring at a 
radius of $1 \farcs 25 $. 
Although the CO arms  contain most ($1.2 \times 10^9$ M$_{\sun}$) 
of the circumnuclear molecular gas, they show only patchy  
H$\alpha$  (Fig. 5) and  Pa$\alpha$ (Maoz et al. 2001) emission. 
In fact, less than 10 \% of the H$\alpha$ flux within the inner kpc 
radius originates in the CO  arms. 
In the western CO arm,  H$\alpha$ emission exists 
from   $11\arcsec$ to $13\arcsec$  and from  $16\arcsec$ to $19\arcsec$, 
in the same regions  where  very faint $K$-band emission is present ($\S$ 4).
It is unlikely that the lack of optically-visible SF along the CO arms 
is  caused only by extinction, since the Pa$\alpha$  and  $K_{\rm s}$ 
images do not reveal significantly more emission.
High resolution  radio continuum  observations, being relatively 
unaffected by dust, would have helped to address the question of 
obscured SF  further. Unfortunately, the only existing 
high resolution ($1\farcs 5$) observations 
do not have enough  sensitivity  to detect any emission 
in the central $40\arcsec$   (J. Wrobel \& J. Condon; private communication). 
Thus, we have to consider the possibility that SF 
may be  truly inhibited in the CO arms. 
One factor leading to this inhibition may be the 
shear induced by the  non-circular kinematics in the  
CO arms ($\S$ 5). The situation is reminiscent of  the inner few kpc 
of M100  where  two massive trailing CO arms with   strong 
streaming motions (e.g.,  Sakamoto et al. 1995)  host a  
large amount  ($1.9 \times 10^9$ $M_{\sun}$) of molecular gas, but have  
comparatively little star  formation, below 
1 $M_{\sun}$  yr$^{-1}$  (e.g., Knapen et al. 1995b).  
A low SFR per unit mass of molecular gas  is also seen 
in the extended gas with non-circular kinematics along the bar 
in NGC~4569 (Jogee 1999) and NGC~7479 (Laine et al. 1999b). 
Another possibility is that the local gas surface density 
in the CO arms of NGC~5248, though   high 
($500-1500~M_{\tiny \sun}$ pc$^{-2}$), is still below the 
critical density required to trigger widespread SF  along 
the arms (see $\S$ 7.2  for a dynamical argument related to this point).

A more detailed look at the  properties of the H{\sc ii} regions and 
SSCs in relation to the molecular  gas shows an interesting evolutionary  
effect.
Table~5 lists several  properties of emission-line complexes   in NGC 
5248, as  derived by Maoz et al. (2001). 
For simplicity, we adopt the same  identification numbers as used 
by these authors.  
The H{\sc ii} regions are listed in Table 5 in order  of decreasing 
extinction, as judged by the  H$\alpha$+[N{\sc ii}]/Pa$\alpha$ flux ratio. 
A value of $\sim$ 10 corresponds  to a case  B recombination
and no reddening,   while 
values below 1 correspond to 3--4 magnitudes of  foreground dust extinction. 
Most H{\sc ii} regions in the starburst ring of NGC~5248 have intermediate 
values of 2--5, corresponding to  an extinction of 1--2 magnitudes. 
The equivalent width of the Pa$\alpha$ line 
(\it f \rm (Pa$\alpha$)/\it f$_\lambda$ \rm  (1.6
$\micron$)), listed in column 6, is  a  measure of the ratio of O stars to K
and M giants and supergiants.
It is of order 400 \AA\ in the first few Myr  
and  falls  by 2 orders of magnitude within 10 Myr (Leitherer et al. 1999; 
Maoz et al. 2001)  under the assumption of a single burst model
whose SF rate (SFR)  decays  exponentially on a timescale of 1 Myr.
While the Pa$\alpha$ equivalent width cannot be used as a detailed  
tracer of age,  it is very likely that complexes  with very large 
Pa$\alpha$ equivalent widths  of 349--943~\AA (such as  3, 4, 16, and 9) 
are  on average younger than complexes with equivalent  widths  
below 50~\AA\  (e.g., 7, 19, 6, and 8).

The four emission-line complexes  (3, 4, 16, and 9) with the largest 
extinction 
turn out to have large  Pa$\alpha$ equivalent widths (361--943 \AA) 
indicative of a young stellar population a few  Myr old, 
but they do not show any  optically-visible continuum  sources (Fig.~6).
This suggests that  the young stellar populations associated with 
these H{\sc ii} regions are  obscured at optical wavelengths by  gas and 
dust. In strong support of this suggestion, we find that 
these  H{\sc ii} regions are located  in the CO arms near bright 
CO peaks  having 
local gas surface densities  of 600--1500 $M_{\tiny \sun}$ pc$^{-2}$. 
A similar avoidance between  line-emitting regions and continuum  
sources has also been noted in the Antennae (Whitmore et al. 1999), 
NGC 2903  (Alonso-Herrero, Ryder, \& Knapen 2001), and several 
 starbursts (Buta et al. 2000). 
NGC~5248 also hosts many H{\sc ii} regions (\# 8, 14, 21, 12, 6, 17, 19, 
7, 18, 20) with H$\alpha$+[N{\sc ii}]/Pa$\alpha$ flux ratios $\ge$ 3 and 
small Pa$\alpha$ equivalent  widths (25--90 \AA). 
These H{\sc ii} regions  (Fig. 6) lie near UV-bright  stellar clusters 
(Fig.~7), most of which  have moderate extinction (0 $<$ A$_{V}$ $<$ 1 mag), 
masses of several 10$^3$ to 10$^5$ $M_{\sun}$, and ages typically 
in the range 10--40 Myr (Maoz et al. 2001). 
The ionized gas  is often distributed in the form of
shells and bubbles which surround the  young  stellar clusters, 
similar to what  is observed in  NGC~1569 (Hunter et al. 2000), 
NGC~5253 (Strickland \& Stevens 1999), and NGC~4214 (Ma\'iz--Apell\'aniz  
et al. 1999). This suggests that stellar winds and  shocks from supernovae  
clear  out the gas and dust. In support of this scenario, we find that in 
NGC 5248 most of these moderately extincted  H{\sc ii} regions  and 
optically-visible  SSCs  are associated with low levels of CO emission 
(Figs. 6 and 7). Furthermore, we find that   the molecular gas
around the brightest complex of H{\sc ii} regions (\# 14) has a bubble
morphology   suggestive of an evacuated cavity (Fig.~6). 
Taken together, the distributions of molecular gas, ionized gas, and 
stellar clusters in NGC~5248 evoke a  scenario where  young 
star-forming regions form within dense gas complexes, and subsequent 
stellar winds and supernovae efficiently clear out the 
gas on timescales less than a few million years. 
\rm

The origin of the CO and dust spirals will be discussed in 
$\S$ 7 where we shall demonstrate that they lie well inside the 
outer ILR (OILR) of the bar.
We discuss here the gas inflow rate from the CO arms 
into the star-forming ring  at $5\arcsec$. 
The net inflow rate  depends on  the mass density profile and 
the vector sum of different velocity components at each radius. 
From a kinematic standpoint, the curvatures in the isovelocity contours, 
the local radial  inflow streaming motions  of 25--40 km~s$^{-1}$ 
along the kinematic minor axis, and the non-circular 
azimuthal streaming motions between $12\arcsec$ to $18\arcsec$ 
along the major axis ($\S 5$) amply demonstrate that 
strong non-circular streaming motions  exist in the CO arms.
If one assumes an average mass density of 600  $M_{\sun}$ pc$^{-2}$, 
a local radial velocity of 15  km~s$^{-1}$, and a width of 200 pc for 
each  CO arm,  the implied gas mass inflow rate along the CO arms 
is a few $M_{\sun}$ yr$^{-1}$. 
This estimate provides an upper limit to the mass inflow rate 
from the CO arms into the star-forming ring  because 
the gas flow is 
expected to diverge and form a `spray-type' flow  (see $\S$ 7.2)  
after the CO arms cross the bar minor  axis and approach 
the star-forming ring.
As a result of the  `spray-type' flow, a significant fraction 
of the gas  inflowing in one CO arm can overshoot the ring and 
enter the other arm, taking a longer time to reach the ring. 
The estimated upper limit of  a few $M_{\sun}$ yr$^{-1}$ 
for the  gas mass inflow rate along the CO arms  is comparable 
to the  current SFR  of  1--2  $M_{\sun}$ yr$^{-1}$  in the starburst ring  
($\S$ 6).  Other estimates of the inflow rate can be made 
by calculating the gravitational torque exerted on the gas by the stars 
or by running tailored hydrodynamical simulations. 
These methods also carry large uncertainties as they depend on the exact
shape of the underlying gravitational potential, as well on uncertain
parameters such as the viscosity, bar pattern speed, sound speed, simplified
approximations of multi-phase medium and SF feedback.

\section {Gas dynamics in the central kpc of NGC~5248}

\subsection {Gas response and dynamical resonances}

A barred potential is made up of different families of periodic stellar
orbits, characterized by a (conserved) Jacobi energy, $E_{\rm J}$, a combination
of energy and angular momentum (e.g., Binney \& Tremaine 1987). The most
important families are those aligned with the bar major axis (so-called $x_1$
orbits) or with its minor axis ($x_2$ orbits) (Contopoulos \&
Papayannopoulos 1980). The $x_1$   family extends between the center and the
bar's corotation radius.  The $x_2$ family appears  between the center
and the ILR if a single ILR  exists, and between the inner and outer
ILRs, if two ILRs exist. When $x_2$ orbits form, they are preferentially
populated at each $E_{\rm J}$. Hence, the orientation of populated periodic
orbits changes by $\pi/2$ at each  resonance.                          

The abrupt change in orientation  at each  resonance is restricted to
(collisionless) stellar orbits. The gas-populated orbits can change their
orientation only gradually due to shocks induced by the finite gas pressure.
Therefore, the gas response to bar torquing leads to the formation of
large-scale offset shocks and a subsequent gas inflow, which slows down after
crossing a resonance. This results in the formation of nuclear rings, each
associated with a parent ILR. Generally, no one-to-one correspondence is
expected between the existence of star-forming (i.e., blue) rings and ILRs, as
rings can merge (Heller et al. 2001) and/or age, becoming devoid of star
formation (Shlosman 1999; Erwin et al. 2001). Knapen et al. (1995b) and
Shlosman (1996) have investigated locations of nuclear rings with respect to
ILRs, concluding that the outer ILR is devoid of any pronounced SF 
because the associated ring sits well inside this resonance. The explanation
to this phenomenon is closely tied to the rate of  dissipation and the sound
speed in the gas, as well as the bar strength which defines the shape of the
$x_2$ orbits and controls which outermost $x_2$ orbits get populated.
At the same time if an inner ILR (IILR) exists, the star-forming ring is positioned
exactly at this resonance,  due to the reversal of gravitational torques.

The exact locations and even the number of ILRs can be inferred reliably only
from non-linear orbit analysis (e.g., Heller \& Shlosman 1996) based on the
knowledge of the galactic potential. In this paper, we use approximate methods
which rely on  SF  morphology and  isophote fitting. The
first method relies on the properties of nuclear rings discussed above. The
dominant circumnuclear ring at a radius of $5\arcsec$  (375~pc) shows that
NGC~5248 has at least one ILR. Whether this galaxy has a second ILR 
near  $r=1.25\arcsec$, where the second inner  H$\alpha$ ring resides, is
not entirely clear. In principle, the latter ring may not be tied to any
dynamical resonance, but could represent a superbubble driven by a recent
starburst event in the nucleus. However, there is no evidence of a bright
young central cluster in NGC~5248 (D. Maoz 2001; private communication),  and
the bright background light from the bulge makes it hard to identify a faint 
central  source.  Hence, it is plausible that NGC~5248 hosts both an OILR 
and an IILR.

A complementary isophote fitting to the deprojected $R$-band and $K_{\rm s}$
(Fig.~10b) images  also provides a guide  to the location of the OILR. 
The isophotes change  gradually from being  oriented along the bar major axis around 
$\sim$~$115\arcsec$, to being perpendicular to the bar around $35\arcsec$.
This gradual twist rather than an abrupt change in orientation results from
the unusually prominent young spiral arms inside the bar corotation. 
The isophote fits provide some insight into locations of dominant $x_1$ and
$x_2$ families of orbits, oriented  parallel and perpendicular to the bar,
respectively. Those corresponding to the outermost $x_2$ orbits cross
the bar minor axis at  $\sim$~$35\arcsec$  and the bar major axis at
$\sim$~$26\arcsec$. These locations can be taken as approximate positions of
the OILR along the  major  and minor axes of the  bar.  The point P2' in 
Fig. 13a and point P2 in Fig. 13b denote the approximate position of the 
OILR along the bar minor axis. The position of the OILR is further 
confirmed by the azimuthal twist of the stellar  spirals arms by  
$\pi/2$ between the OILR  and the CR, and an additional $\pi/2$ between 
the CR and the OLR (Fig.~1) The latter is estimated to lie around $\sim
230\arcsec$, at the edge of the outer disk where the spiral arms end ($\S$ 3).
Furthermore, Laine et al. (2002) have analyzed a subsample of barred galaxies
among 112 disks, concluding that the ratio of $r_{\rm ILR}/D_{\rm 25}$ is
about 0.06, where $r_{\rm ILR}$  is taken along the bar major axis. For
comparison, this ratio is 0.07 in NGC~5248, assuming  $r_{\rm
ILR}$=$26\arcsec$ and  $D_{\rm 25}$=$370\arcsec$ (Table~1).              

To summarize, we discuss briefly three possibilities concerning the number of
ILRs in NGC~5248: 
\\
\indent  {\it (i). One ILR:} In this case, with the the ILR at around 
$26\arcsec$ (2 kpc),  it is difficult to explain why the star-forming ring 
at 375~pc lies so deep  inside the resonance. NGC~5248, therefore,
would appear as an extreme case. Another limitation of this scenario is that it
does not provide an explanation for the inner H$\alpha$ ring at 95~pc.
\\    
\indent  {\it (ii). Two ILRs:} In this case, the OILR is at 2~kpc 
and the IILR at 95~pc. This option is attractive because as 
required by numerical simulations,  the OILR  itself is devoid of pronounced  SF,  
there is an  inner star-forming ring  at the IILR, and an  
outer ring  at 375~pc  between the ILRs. However, one 
discrepancy is that in simulations, the IILR ring  generally has more 
prominent  SF than the outer ring, while the converse would appear to hold 
here. 
\\
\indent {\it (iii). Three ILRs:} In this case,  the OILR is at 2~kpc, the IILR at
375~pc, and the innermost resonance, hereafter OILR$_{\rm 2}$, is 
at $\sim$ 200~pc, outside the ring at 95~pc. 
This option sets the more prominent ring with active SF, namely the 375~pc ring,  
at the IILR.   A combination of high-resolution
velocity fields and future detailed modeling can help distinguish between the 
different possibilities.  Meanwhile, for brevity, we shall work within the
framework of 2 to 3 ILRs, as these models avoid most of the controversies.     
     
Aside from the morphology of H$\alpha$ rings, kinematic properties can 
in principle  constrain the number and location of ILRs. For a weak bar where
the linear epicyclic approximation is (very roughly) tolerable, we can estimate
the location of ILRs if we have a reliable measure of the angular speed
$\Omega$ and the epicyclic frequency $\kappa$. The latter quantity depends
sensitively on the  \it circular \rm  velocity as a function of radius.  
The rotation curve (Fig.~12) for NGC~5248 suffers from non-circular motions
between $8\arcsec$ and $11\arcsec$, and $12\arcsec$ and $18\arcsec$ 
($\S$ 5). In consequence, the linear  resonance diagram is too noisy to
provide meaningful estimates of the location of the ILRs. However, an upper
limit to the bar pattern speed  $\Omega_{\rm p}$  can be inferred from it. 
$\Omega$ is  of order 30~km~s$^{-1}$ kpc$^{-1}$ around a radius of $70\arcsec$, 
and for a flat rotation curve it must be even lower  at the CR of
$115\arcsec$ (Paper~I and $\S$ 3). The condition [$\Omega$~=~$\Omega_{\rm
p}$] at CR, therefore, implies   that the bar pattern speed is below
30~km~s$^{-1}$ kpc$^{-1}$. This low pattern speed will be used in 
$\S$ 7.2.

An interesting,  albeit speculative  possibility is that we may be 
observing a time-dependent rather than a quasi-static phenomenon in
the disk of NGC~5248. Numerical modeling  of the evolution, geometry, and
associated SF of circumnuclear rings shows  that when gas
inflow rates along  the bar are large, the ring  is typically oval 
in shape and leads the bar by $~\sim 50^\circ-60^\circ$ (Knapen et al.
1995b; Shlosman 1996). Once  gas inflow rates slow down substantially, the  
circumnuclear ring acquires  a round shape and a relatively uniform
distribution of SF, typically  after a characteristic timescale
of order  $10^8$~yrs.  In the  deprojected H$\alpha$ image of NGC~5248, the
star-forming  ring at 375~pc has such  a quasi-circular appearance, with star 
formation uniformly distributed azimuthally. Furthermore, a significant
fraction of the stellar clusters in NGC~5248  have ages of order 100 Myr (Maoz
et al. 2001), consistent  with the above characteristic evolutionary 
timescale.  On the other hand, the massive CO arms with large non-circular 
motions joining the ring, the young age (10--20~Myr; Maoz et al. 2001) 
of many SSCs in the ring, and the young age (10 Myr; $\S$ 4) of the 
$K$-band arms which delineate them, all point to a recent triggering of 
SF in the last 10--20~Myr. Taken together, the complicated 
morphology of the stars and ISM in the vicinity of  ILRs,  and the range of
ages seen in the  stellar clusters allow for a strongly time-dependent
evolution  of the inner kpc.

\subsection {Bar-driven gas density waves inside the ILR}

The present work provides ample observational evidence in favor of 
continuity of grand-design spiral structure from the nuclear (sub-kpc) to 
10~kpc scales in the barred galaxy NGC~5248. In particular, our observations
($\S$ 3) show that NGC~5248 hosts spiral structure in stars, dust, and gas 
over a striking two orders of magnitude in spatial scales. They also
demonstrate that spiral structure in NGC~5248 winds up over a large
azimuthal angle, at least $\pi$, deep inside the OILR ($\S$ 7.1). 
The two massive CO and dust spirals (Fig.~4b) extend between $20\arcsec$ and
$5\arcsec$, connecting to the inner young  $K$-band arms ($\S$ 3; Fig.~8) 
which continue from $8\arcsec $ to $3\arcsec$, delineating  the SSCs (Fig.~7) 
in the $5\arcsec$ star-forming ring.  Still closer to the center, 
between  $1\arcsec $ and $4\arcsec$, reside the nuclear dust spiral, the
double-peaked molecular feature, and the nuclear H$\alpha$ ring ($\S$ 3). 
To account for the observed properties, we next compare NGC~5248 to 
theoretical models of gas response deep inside the ILR of a barred 
potential.

According to the density wave theory (e.g., Lin \& Shu 1964; Lin, Yuan, \&
Shu 1969; Bertin et al. 1989a,b), \it stellar \rm  density waves
cannot propagate inwards across the ILR and are damped at smaller radii.
However, \it gaseous \rm density waves, in principle, can exist at most radii,
depending on the degree of self-gravity in the gas. If the gas self-gravity is
ignored,  the gaseous wave propagation is limited to the region between the
ILRs (for the case of two ILRs), or  to the region between the center and the
ILR (for the case of one ILR). The propagation of gaseous density waves in 
the central regions of disk galaxies has attracted  recent attention 
and has been investigated with hydrodynamical simulations  
(Englmaier \& Shlosman 2000). 
In these simulations, strong bar shocks which exist in the vicinity of
the ILR(s) trigger gas density waves which propagate radially inwards and
are sheared by galactic rotation. The transition between the gaseous wave
driven by the outer shocks and the lower amplitude wave which is
propagating inwards happens near the transition radius $R_{\rm t}$, where
shocks cross the bar major axis (see the point P4' in Fig.~13c).  The simple
explanation for the sudden drop in wave amplitude seems to be the existence
of a `spray-type' flow in this region. In other words, the flow diverges here
suddenly due to trajectory curvature and the surface density drops as a
result. This type of flow happens only when  the wave  approaches the major
axis of the bar, while  at other azimuths,  the flow streamlines converge.

Englmaier \& Shlosman (2000) ignored the contribution of self-gravity 
in their dispersion relationship for gaseous waves deep inside the ILRs
of barred galaxies and in their modeling. This approach was motivated by the 
observed low arm-interarm contrast of $\sim 2$ and the lack of star
formation along  the nuclear dust spiral in NGC~5248, observed in $J-K$ 
between $1\arcsec$ and $4\arcsec$ (Laine et al. 1999).  The shape of the
resulting gaseous spiral depends on two parameters: the degree of central
concentration and the sound speed in the gas. One can obtain a range of
shapes, from a tightly-wound to an open one. The tightly-wound spirals in fact
degenerate into a featureless disk. The theoretical and numerically-modeled
shapes of the spirals and  their amplitudes are found to agree well within
the transition radius $R_t$. Outside this radius  the theoretical 
curve suffers from a number of simplifications, most notably the neglect
of the gas self-gravity term.  This  leads to a singularity in the analytical
expression for the pitch angle, tan~$i_p \sim 1/R|k|$, at the ILR, where
the wavevector $k\rightarrow 0$. This unphysical behavior can be corrected
by adding the gravitational  term, $-2\pi G \Sigma |k|$, where $\Sigma$ is the
gas surface density, to the dispersion relationship (Eq.~1 in Englmaier \&
Shlosman). The prime motivation for introducing this correction in this 
paper is the existence of massive molecular (CO) arms with a large
(arm/interarm) contrast in NGC~5248. 

Figs.~13a and 13c show  the model gas response, respectively, along the 
entire bar and inside the OILR. The corresponding observed spirals are 
shown, respectively,  in Figs. 13b and 13d. The massive CO arms 
in NGC~5248 appear to 
correspond to the high amplitude non-linear gaseous waves propagating 
inwards  across the OILR, in which case they would have to lie outside the
transition radius $R_{\rm t}$. On the other hand, the double-peaked CO
feature F1 and nuclear dust spiral, which lie  interior to the starburst ring
of $5\arcsec$ radius, appear to be associated with the linear wave. 
This suggests that $R_{\rm t}$~$\sim$~$5\arcsec$--$6\arcsec$. Note that the
minimum of the spiral pitch angle $i_p$ (Fig.~14a) lies near this estimated
$R_{\rm t}$. 

Fig.~14a displays the observed (dotted line) pitch angles of gaseous spirals.
The pitch angle of the dusty spiral is marked with a dot at $5\arcsec$. The
solid line is a smooth eyeball-fit to the observed pitch angle, which we used
below for the self-gravitating case.  In the case where the gaseous density
wave and large-scale bar have  the same pattern speed, and when the 
self-gravity of the gas can be neglected, the model of Englmaier \& Shlosman (2000) 
predicts a unique run of pitch angle with radius, once the sound speed, the pattern
speed, and the rotation curve are specified. As the non-linear gaseous spiral
wave travels inside the OILR, the model predicts that its  pitch angle will
drop until it gets into resonance with  the linear gaseous wave at $R_{\rm
t}$. Interior to $R_{\rm t}$, the pitch angle is expected to increase.  
On the basis of this predicted behavior, Fig.~14a suggests that $R_{\rm t}$
lies around $5\arcsec-6\arcsec$.  One can discuss separately the behavior of
two branches, for the radii smaller or larger than $R_t$. The left branch is
formed by the dust nuclear spiral, while the right one is associated with the
dusty spiral which follows the CO arms. Note that the pitch angle of the two
trailing  CO arms can be reliably measured  only from a radius of $18\arcsec$
(1.3~kpc) to $6\arcsec$  (450~pc). From $6\arcsec$ and $3\arcsec$, the CO
pitch angle cannot be reliably measured as the emission is too weak. 
\rm 
This region between  $3\arcsec$ and $6\arcsec$  is where the spiral
structure appears to continue in the form of $K$-band arms which delineate 
the SSCs in the starburst ring. As discussed in $\S$ 6, the weak CO emission 
in this region is likely due to CO near the SSCs being  cleared out by winds
and  supernovae, or dissociated.  Fortunately, the pitch angle  of the 
nuclear dust spiral (Laine et al. 2001)  provides empirical data 
between $1\arcsec$ and $4\arcsec$.   
%
\rm
 
One can also infer the physical characteristics of gaseous spirals 
which result from the smooth eyeball-fit to the observed pitch angle (Fig.~14a;
solid line) and the self-gravity term. These inferred characteristics include 
the Toomre  $Q$ ($\equiv v_s\kappa/\pi G \Sigma$) parameter, 
and the gas surface density.  Here,  $v_s$
is the gas sound speed and $\kappa$ is the epicyclic frequency. Fig.~14b 
displays the effective $Q$ for $v_s=15$~km~s$^{-1}$ and the bar pattern speed
$\Omega_{\rm p}=25$~km~s$^{-1}$~kpc$^{-1}$. This bar pattern speed 
is justified in $\S$ 7.1.  The  $Q$-curve tends asymptotically
to infinity on both sides of the `gap' between $6.5\arcsec-12\arcsec$. Larger
values of $Q$ correspond to progressively less `self-gravity.' This gap has a
counterpart in the surface gas density profile (Fig.~14c), where
$\Sigma\rightarrow 0$. The gap in the above analysis displays the effect of the
$5\arcsec$ ring and the associated $K$-band arms extending to about
$8\arcsec$.  One should not forget also that the non-circular motions are 
large  in this region adjacent to the ring ($\S$ 5).  Clearly, 
neglecting the contribution of the stellar component is not permitted here. 
  
Our estimate for the position of $R_{\rm t}$ in Fig.~14a can be independently
tested. Namely, close to the OILR, large deviations from purely circular 
motions are expected (Fig.~15a). On the other hand, inside $R_{\rm t}$, the 
model predicts an azimuthal flow speed which is close to the circular
speed. Thus, around $R_{\rm t}$, the observed and circular speed  are expected
to converge (Fig.~15a). Figs.~15b and 15c show the observed difference between 
the circular speed curve and the CO velocity. The largest deviation from
circular motions occurs between a radius of  $7\arcsec$ and $10\arcsec$ and a
sharp change is seen at about $6\arcsec$, where the observed and
circular speeds converge. This supports our estimate of $R_t\sim 6\arcsec$.
The kinematical properties are  therefore  consistent with the model
predictions.

The  above combination of multiwavelength observations and modeling
suggests that when the right conditions are met, a bar can drive 
a gaseous SDW which winds over a large azimuthal angle and extends
deep inside the ILR. The perturbations thus generated inside the OILR can
be  \it strong and self-gravitating,  \rm such as the massive 
CO arms in NGC~5248. Overall, the  spiral structure induced by the bar
may lead to  an appreciable gas inflow rate relevant for  the fueling of  
starburst  activity in nuclear rings --- a possibility  strongly hinted 
at in NGC 5248 by the large non-circular motions in the CO arms 
and the massive recent SF along the young $K$-band spirals ($\S$ 6).  
A relatively low  central mass concentration 
(evident through a  shallow rotation  curve) which prevents the spirals from
winding up, is a crucial condition favorable to the propagation of the 
gaseous SDW deep inside the ILR.  
This condition is more easily met in late-type than in early-type  spirals.  
Conversely, in the presence of a large central 
mass  concentration, the  models of Englmaier \& Shlosman (2000)  show 
that the gaseous wave driven by the outer bar shocks can die away 
rapidly inside the ILR. This very condition, interestingly enough, leads 
to the formation of ILRs and favors the formation and dynamical decoupling of
secondary nuclear bars by causing the inner regions of a bar to rotate much
faster than primary bar pattern  speed. 
We note that the rotation curve of NGC~5248 is shallower than that  of 
many well-studied barred  galaxies  such as NGC~6951, NGC~4102 (Jogee 1999),  
NGC 2782 (Jogee et al. 1999), and  M100 (Knapen et al. 1995a).  
In NGC~5248,  the offset spiral dust lanes  and CO arms 
cover at least $180^\circ$  in   azimuth inside the OILR before they join a  
starburst ring (Fig. 6 and Fig. 13d), while in  
in NGC~6951  (Jogee 1999) and  M100 (Knapen et al. 1995a), 
the corresponding azimuthal  angle is  only $\sim$~$90^\circ$.
In NGC~2782 which hosts a nuclear bar,  
the CO rotation curve  has a turnover velocity of above 300 ~km~s$^{-1}$, 
and is much steeper  than in NGC~5248  (Jogee et al. 1999).

\section{Summary and Discussion} 

We present a study of the grand-design spiral structure, gas dynamics, 
and circumnuclear SF in the nearby barred galaxy NGC~5248, 
based on a multi-wavelength dataset and  theoretical modeling 
of gas dynamics. The extensive dataset includes high ($1 \farcs 9  \times 1
\farcs 4$) resolution OVRO  CO (1--0) observations, optical and NIR
ground-based data, along with  {\it HST}  images and  Fabry--Perot H$\alpha$
observations. This study provides  the best evidence to date for a 
strong dynamical coupling between the nuclear (sub-kpc) region and the
large-scale features   in the outer galactic disk.   It shows how bar-driven
spiral structure in NGC~5248 extends over  two orders  of magnitude in
spatial scales, shapes disk evolution, and fuels  SF on
progressively  smaller scales.

The grand-design spiral structure is particularly evident in  the dust
spirals  which cover at least 360$^\circ$  in azimuth and continue from a
radius of 70~kpc  down to 375~pc, penetrating  the OILR of the bar located  
at  2.0~kpc.  Interior to the OILR,  two massive trailing molecular spiral
arms cover nearly $180^\circ$ in azimuth  from  a  radius  of 1.5~kpc  to 
375~pc,  where they feed  a  circumnuclear  starburst ring. The molecular 
spirals  are massive, containing  $1.2 \times 10^9$ M$_{\sun}$  of gas, and 
show non-circular streaming  motions of  20 to 40   km~s$^{-1}$. At a radius
of $\sim$~600~pc, they  connect to two $K$-band arms which  cross the
starburst ring and  continue down  to 225~pc. The UV-bright SSCs seen 
in the \it HST \rm image lie along these   $K$-band spirals. The particularly narrow 
width ($\sim$~110~pc) of the $K$-band arms  suggests  a  young  dynamical age 
of $\sim$~10~Myr,   comparable to the age of the SSCs (10--40~Myr; Maoz et al 2001) 
and that of the  supergiants (8--10~Myr)  which often dominate
$K$-band emission in starburst regions. Altogether, the data suggest that 
the $K$-band arms are young,  and recent  SF in the ring  at
375~pc has been triggered  by a bar-driven density wave.  The density wave
may have even propagated  into the  central 100~pc since the  $K$-band arms
appear to connect  to a  grand-design nuclear dust spiral which continues 
inwards from $\sim$~225 pc to  $\sim$~75 pc (Laine et al. 1999). This dust
spiral crosses a  second H$\alpha$ ring of radius  95~pc and a double-peaked
molecular feature.

We estimate a SFR of about 2 $M_{\sun}$ yr$^{-1}$ 
in  the inner few kpc of  NGC~5248. Only a small fraction of this 
SFR originates in the massive CO spiral arms although they contain most of
the  molecular gas. 
The shear induced by the non-circular kinematics, as well as sub-critical
gas densities,  may be responsible for the low SF efficiency in
the arms. The latter point is supported by modeling. 
We also find that the four  emission-line  complexes which  have  the 
largest extinction and line ratios consistent with an embedded, 
optically-invisible, young  stellar population,
a few Myr old,  are located in the massive CO arms, near bright 
CO peaks with  large gas surface densities of 600--1500~$M_{\tiny \sun}$ pc$^{-2}$.  
Conversely, visible SSCs which have  moderate extinction  (0 $<$ $A_V$ $<$ 1 mag) 
lie in the starburst ring where they are  associated with low levels of CO emission  
and are surrounded by shells  of ionized gas. Overall, the distribution of 
molecular gas, ionized gas, and stellar clusters in NGC~5248 are consistent 
with a picture  where  young star-forming regions are born within dense 
gas complexes and subsequent stellar winds and supernovae efficiently clear 
out gas on timescales less than a few million years.  

There is mounting evidence from recent high resolution (10--100~pc)
ground-based and \it HST \rm observations  that grand-design dust and  gas
spirals are common in  the inner kpc of  galaxies. In NGC~5248, a bar-driven 
SDW has led  to massive CO spiral arms which cover a large range in radius 
inside  the OILR and feed a starburst ring. 
To account for such massive perturbations we have generalized the 
Englmaier \& Shlosman (2000) models of gaseous SDW by incorporating the
effect of gas self-gravity. We thereafter find good agreement between the
modeled and  observed gas morphology, gas kinematics, and pitch angle of the 
spirals. This  suggests that grand-design spirals in the inner kpc can 
be explained by the propagation of gaseous density waves deep inside the
ILR of the barred potential, when the right dynamical conditions are met.
In particular, our study confirms that a low central mass concentration 
(evident through a shallow rotation curve),  which may be common in many 
late-type galaxies, is particularly favorable to  the propagation of
bar-driven  gaseous density waves deep into  the central region of the galaxy. 
Conversely, a large central mass concentration favors other processes such
as the formation and decoupling of nuclear bars.

\acknowledgments
Support for this work was generously provided by a grant from the K.T. and E.L. 
Norris Foundation, NSF grant AST 99-81546, and an AAUWEF Fellowship. I.S.
acknowledges support under NAG 5-10823, HST GO-08123.01-97A and WKU
516140-02-07. We thank Dan  Maoz, Aaron Barth, and Alex Filippenko 
for generously providing the {\it HST} continuum subtracted H$\alpha$+[N{\sc
ii}] images.  We thank Bruce Elmegreen, Chi Yuan,  and  Frank Shu for 
interesting discussions. This publication is  based on observations 
made with the NASA/ESA Hubble Space Telescope, obtained from the Data Archive 
at the Space Telescope Science Institute, which is operated by the
Association of Universities for Research in Astronomy, Inc., under NASA
contract NAS 5-26555.  
The Isaac Newton and William Herschel Telescopes are operated on the
island of La Palma by the Isaac Newton Group (ING) in the Spanish
Observatorio del Roque de los Muchachos of the Instituto de Astrof\'\i
sica de Canarias. Data were partly retrieved from the ING archive.


\clearpage

\begin{deluxetable}{ll}
\tabletypesize{\scriptsize}
\tablecaption{Adopted Parameters for NGC 5248. \label{tbl-1}}
\tablewidth{0pt}
\tablehead{  
}
\startdata
Hubble type  &   SAB(rs)bc$^{b,c}$ \\
Nuclear type & H{\sc ii}{2}$^{d,e}$ \\
$D_{\rm\tiny 25}$ &  $370\arcsec$$^{b}$ \\
Distance \it D \rm &  15.3$^{b}$  Mpc \\
Inclination \it i \rm & 40$^{c}$ $\pm$ 4 $\deg$\\
Line of nodes & 105$^{c}$  $\pm$ 2  $^\circ$ \\
Center(B1950.0)  & R.A.= 13h 35m 02.54s$^{b}$ \\
                 & Dec = $09^\circ$ $08\arcmin$ $21.56\arcsec$$^{b}$ \\
$V_{\rm sys}$  & 1153$^{b}$  km~s$^{-1}$ \\
\enddata 
\tablerefs{
a. Nearby Galaxies (NBG) Catalogue (Tully 1988); 
b. RC3 (de Vaucouleurs et al. 1991); 
c. Jogee et al. 2002 
d. Ho, L.~C., Filippenko, A.~V., \& Sargent, W.~L.~W. 1997; 
e. Kennicutt, Keel, \& Blaha 1989.
}
\end{deluxetable}

\clearpage

\begin{deluxetable}{ll}
\tabletypesize{\scriptsize}
\tablecaption{Parameters of OVRO  channel maps. \label{tbl-2}}
\tablewidth{0pt}
\tablehead{  
}
\startdata 
Pointing center (B1950.0) & R.A.= 13h 35m 02.60s  \\
                         & Dec = $09^\circ$ $08\arcmin$ $22.0\arcsec$ \\
Projected baselines &  12 to 483 m  \\
Transition  &   CO (1--0) at 115 GHz \\
Half power beam width   &   $65\arcsec$  \\
Spectral bandwidth & 240 MHz, 600 km~s$^{-1}$ \\
Spectral resolution &  5.2 km~s$^{-1}$ \\
Synthesized beam &   1 \farcs 9 $\times$ 1 \farcs 4, $140 \times 100$ pc \\
                     & PA = $-80.91^\circ$ \\
Emission channels &  1012 to   1293  km~s$^{-1}$ \\
Peak emission   & 210 mJy/beam \\
Peak $T_{\rm b}$$^{a}$   &    6.9 K       \\
Typical r.m.s. &  17 mJy per beam \\
Flux in OVRO map &  385 Jy km~s$^{-1}$ \\
M$_{\rm gas}$$^{b}$   & $1.4 \times 10^9$ M$_{\sun}$ \\ 
$F_{\rm sd}$$^{c}$   &  85 \% \\ 
Single dish flux  &  450$^{d}$  Jy km~s$^{-1}$ 
\enddata 
\tablecomments {
a. $T_{\rm b}$ = Brightness temperature in Rayleigh-Jeans approximation. 
b. $M_{\rm gas}$ is the mass of molecular gas including the contribution 
of helium, assuming  a standard Galactic CO-H$_{\rm 2}$ conversion factor 
and a  solar composition (see text). 
c.$F_{\rm sd}$ is the fraction of single dish flux recovered.
d. From Five College Radio Observatory observations (Young et al. 1995)
}
\end{deluxetable}

\clearpage

\begin{deluxetable}{ccccc}
\tabletypesize{\scriptsize}
\tablecaption{Summary of data on NGC 5248 presented in this paper.\label{tbl-3}}
\tablewidth{0pt}
\tablehead{
\colhead {Data }&
\colhead {Instrument/telescope }&
\colhead {$t_{\rm exp}$}&
\colhead {Field of view}&
\colhead {Source} \\
\colhead {(1)}&
\colhead {(2)}&
\colhead {(3)}&
\colhead {(4)}&
\colhead {(5)} \\
}
\startdata 
CO (1--0) & OVRO mm array &  8 x 6 h & $65\arcsec \times 65\arcsec$ & a \\
H$\alpha$+[N{\sc ii}] & WFPC2/{\it HST \rm}  & 4800 s & $35\arcsec$ & b \\
F336W  & WFPC2/{\it HST \rm}  & 2300 s & $35\arcsec$ & b \\
F547M  & WFPC2/{\it HST \rm}  & 900 s & $35\arcsec$ & b \\
F814W  & WFPC2/{\it HST \rm}  & 1100 s & $35\arcsec$ & b \\
Harris $R$ &  WFC/INT 2.5 m  &  30 min &  11\farcm 3 $\times$  22\farcm 5 & a,c \\
Harris $B$ & PFC/INT 2.5 m &  20 min &  $11.0\arcmin \times  11.0\arcmin $ & d \\
$K_{\rm short}$ & INGRID/WHT 4.2 m & 12 min & 5\farcm 1 $\times$  5\farcm 3 & e \\
Fabry-Perot H$\alpha$ & TAURUS II/WHT 4.2 m  & 55$\times$110 s  & $ 1.4\arcmin \times  1.4\arcmin$ & f  \\
$J-K$  map & MONICA-AO/CFHT 3.6  m & 180 s &  12 \farcs 4 $\times$ 12 \farcs 4 & g \\
$J-K$  map & MONICA /CFHT 3.6 m & 300 s &  $1.0\arcmin$ $\times$ $1.0\arcmin$ & g \\
\enddata 
\tablecomments{ Columns are : 
(1) Description of the data.
We quote the transition for the interferometric data and the 
filter for the images; 
(2) Instrument and telescope. Abbrevations used are  
OVRO = Owens Valley Radio Observatory; 
WFPC2/{\it HST \rm} =  Wide Field Planetary Camera 2 on the Hubble  Space Telescope; 
WFC/INT = Wide Field Camera on the Isaac Newton Telescope at 
La Palma; 
PFC/INT = Prime Focus Camera on the Isaac Newton Telescope at 
La Palma; 
INGRID/WHT = INGRID camera on the William Herschel Telescope  at La 
Palma; 
TAURUS II/WHT  = TAURUS II instrument in Fabry Perot mode 
on the  William Herschel Telescope at La Palma;   
MONICA-AO/CFHT = Combination of  Montreal NIR camera MONICA with  the
adaptive optics system PUEO on the  Canada-France-Hawaii Telecope at Mauna
Kea;  
(3) On-source exposure time; 
(4) Field of view.  For OVRO, the primary half power beam width 
at 115 GHz is quoted; 
(5) Source of data.
}
\tablerefs{
a. This work; 
b. Archival {\it {\it HST \rm}}  images  first published by Maoz, Barth, Ho, \& 
Fillipenko  (2001); 
c. Jogee et al. (2002)  
d. Isaac Newton Group  archive; 
e. Knapen et al.,  in preparation;
f. Laine et al. (2001);
g. Laine et al. (1999)
}
\end{deluxetable}

\clearpage

\begin{deluxetable}{ccccccccc}
\tabletypesize{\scriptsize}
\tablecaption{SFR  from different tracers.\label{tbl-5}}
\tablewidth{0pt}
\tablehead{
}
\startdata 
$L$(1.4GHz)$^{a}$  & $3.7 \times 10^{21}$   W Hz$^{-1}$ \\
SFR(1.4 GHz)  &   3.2 M$_{\sun}$ yr$^{-1}$  (global)  \\
$L$(FIR)$^{b}$  &  $8.1 \times 10^{9}$  $L_{\sun}$  \\
SFR(FIR)       &   3.1 M$_{\sun}$ yr$^{-1}$  (global) \\
$L({H\alpha})$$^{c}$ &  $2.3 \times 10^{41}$ erg s$^{-1}$ \\
SFR(H$\alpha$) &   2.1  M$_{\sun}$ yr$^{-1}$ (circumnuclear) \\
SFR($K$-band )$^{d}$ &  1.5 M$_{\sun}$ yr$^{-1}$ (circumnuclear) \\
\enddata 
\tablecomments { The table shows  the derived SFRs  
using different tracers as described in the text. 
The term ``global'' refers to the entire galaxy while the term 
``circumnuclear'' refers to the inner 400 pc radius, including the
two rings of SF. Notes refer to: 
a. Radio continuum luminosity density at 1.4 GHz 
from Condon  et al. (1990);
b. FIR luminosity calculated from the IRAS 60 and 100 $\micron$ 
flux density following Helou  et al. (1988); 
c. Extinction corrected H$\alpha$ luminosity  from {\it HST \rm} WFPC2 data 
(Maoz et al. 2001). 
d. Average SFR based on ages and masses of ``hot spots'' derived from 
 near-infrared colors (Elmegreen et al. 1997).  
}
\end{deluxetable}

\clearpage

\begin{deluxetable}{ccccccccc}
\tabletypesize{\scriptsize}
\tablecaption{Emission-Line Complexes in NGC 5248.\label{tbl-4}}
\tablewidth{0pt}
\tablehead{
\colhead {Id}&
\colhead {$\Delta$ $\alpha$} & 
\colhead {$\Delta$ $\delta$} & 
\colhead {Radius} & 
\colhead {$\frac{\it f \rm (H\alpha+[N{\sc ii}])}{\it f \rm (Pa\alpha)}$  } & 
\colhead {$\frac {\it f \rm (Pa\alpha)} {\it f_\lambda \rm  (1.6 \micron)}$ } &
\colhead {\it f \rm (H$\alpha$)} & 
\colhead  {\it f \rm (Pa$\alpha$)} \\
\colhead {(1)} &
\colhead {(2)} &
\colhead {(3)} &
\colhead {(4)} &
\colhead {(5)} & 
\colhead {(6)} & 
\colhead {(7)} & 
\colhead {(8)} \\
}
\startdata 
3  & 
5.19  & 4.24  & 0.18   & 0.2  & 361 & 1.2   & 4.9     \\
4  & 
5.73  & 4.03  & 0.18   & 0.3  & 515 & 1.9   & 6.8     \\
16 & 
-6.70 & -2.75 & 0.36   & 0.4  & 566 & 8.8   & 21.4      \\
9  & 
10.18 & -0.5  & 0.18   & 1.1  & 943 & 7.8   & 6.8      \\
11 & 
5.50  & -2.69 & 0.27   & 1.5  & 89  & 17.5  & 11.7      \\
13 & 
0.98  & -6.24 & 0.32   & 1.6  & 170 & 21.8  & 13.6      \\
15 & 
-5.01 & -2.31 & 0.36   & 1.7  & 121 & 33.8  & 20.4      \\
2  & 
2.97  & 2.99  & 0.18   & 1.9  & 149 & 16.9  & 8.8      \\
10 & 
6.84  & -1.72 & 0.18   & 2.5  & 146 & 12.1  & 4.9     \\ 
5  & 
6.04  & 3.21  & 0.18   & 2.5  & 349 & 17.0  & 6.8      \\ 
1  & 
3.25  & 4.61  & 0.32   & 2.9  & 111 & 30.6  & 10.7      \\
18 &
-3.74 & 2.74 &  0.23   & 3.2  & 58  & 21.8 & 6.8     \\
20 & 
-2.37 & 7.44 &  0.18   & 3.5  & 186 & 13.6 & 3.9      \\
7  & 
4.52  & 0.62  & 0.32   & 3.7  & 41  & 35.8  & 9.7      \\
19 &
-3.28 & 3.79 &  0.23   & 3.8  & 45  & 18.7 & 4.9      \\
17 &
-6.21 & 0.93 &  0.36   & 3.9  & 58  & 45.2 & 11.7      \\
6  &
3.65  & 0.96  & 0.27   & 4.0  & 47  & 35.2  & 8.8      \\ 
12 &
4.10  & -3.90 & 0.45   & 4.1  & 87  & 104.7 & 25.3      \\
21 &
-7.89 & 4.16 &  0.18   &  4.9  & 80  & 33.6 & 6.8      \\   
14 &
-2.32 & -3.93 & 0.91   & 5.0  & 61  & 358.0 & 72.0      \\ 
8  & 
4.50  & -0.54 & 0.23   & 7.0  & 27  & 27.1  & 3.9      \\ 
\enddata 
\tablecomments { 
The emission line complexes are listed in order of decreasing extinction, 
as measured by  the  H$\alpha$+[N{\sc ii}]/Pa$\alpha$ flux ratio 
(Maoz et al. 2001). Columns are :
(1) Identification number for the emission-line complex  used by 
 Maoz et al. (2001); 
(2)-(3)  Offset along the R.A. and declination direction in arcseconds 
from the galaxy nucleus whose pixel coordinates are  (422.16,396.02); 
(4) Angular radius in arcseconds of the circular aperture used for 
flux measurements; 
(5) The  H$\alpha$+[N{\sc ii}]/Pa$\alpha$ flux ratio which is a measure 
of extinction; 
(6)  The equivalent width of the Pa$\alpha$ line in \AA, 
defined as the flux ratio of the Pa$\alpha$ line and the  continuum in the 
1.6 $\micron$ band. This is an  age indicator for a single burst model. 
(7)-(8)  The  H$\alpha$+[N{\sc ii}] and  Pa$\alpha$ line fluxes 
in units of 10$^{-16}$  ergs s$^{-1}$ cm$^{-2}$   without any correction 
for extinction.  At the distance of 15.3 Mpc for NGC 5248, 
10$^{-16}$  ergs s$^{-1}$ cm$^{-2}$  corresponds to a luminosity of 
$2.8 \times 10^{36}$ ergs  s$^{-1}$.
}
\end{deluxetable}

\clearpage

\clearpage
\begin{figure}
\plotone{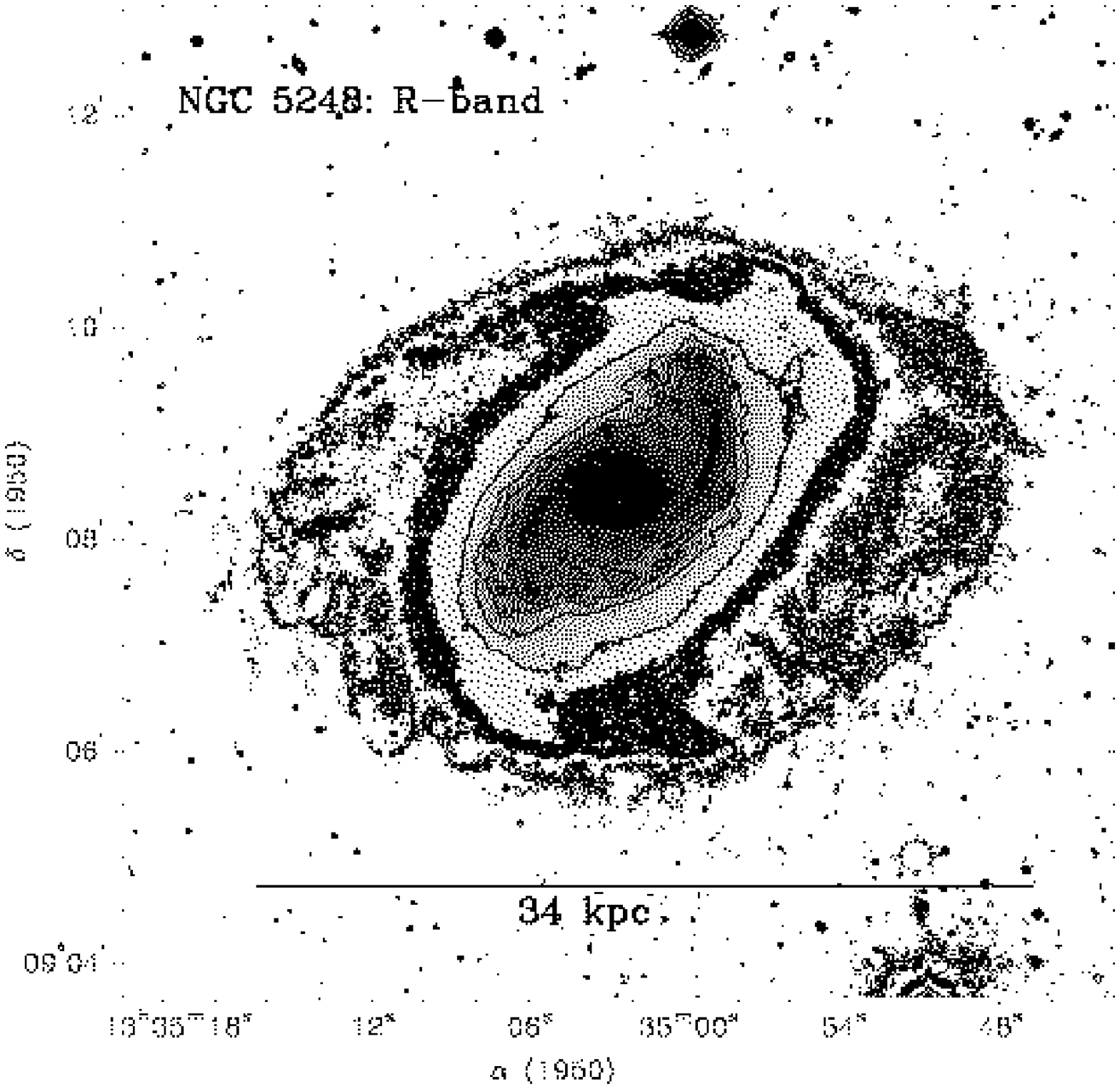}
\caption{
\bf  The large-scale stellar bar:   
\rm 
The  deep wide field  $R$-band  image from  Jogee et al. (2002) 
reveals a  hitherto unknown  prominent  stellar 
bar  with a semi-major axis  of $\sim 95\arcsec$ (6.4~kpc). The bar is
embedded within a faint more circular   outer disk which is  visible 
out to a radius of  $230\arcsec$ (17.2~kpc). The bright inner arms 
of the grand-design spirals  lie  on the leading edge of the 
stellar bar, delineating SF. 
\bf
[For or a version of the paper with high resolution figures see 
ftp://ftp.astro.caltech.edu/users/sj/astroph/n5248-p2-highres.ps.gz]
\rm
\label{fig1}}
\end{figure}

\clearpage
\begin{figure}
\plotone{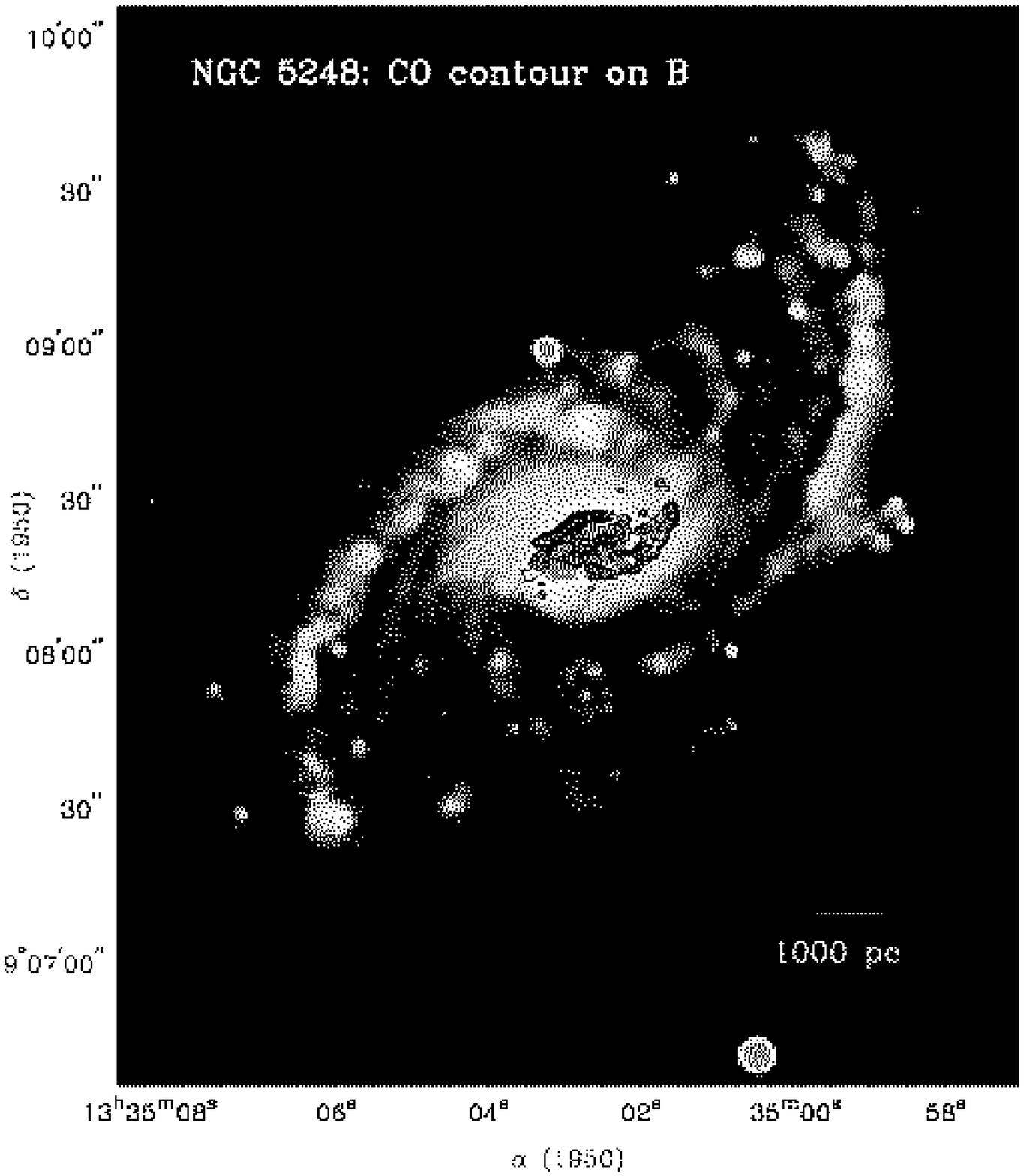}
\caption{
\bf The optical grand-design spirals:   
\rm
The CO (1--0) intensity map  (contours) of the central $40\arcsec$
is overlaid  on a $B$-band  image. Two bright  
stellar spiral arms, lined with young star-forming regions, are 
conspicuous between $\sim 30\arcsec$ and  $90\arcsec$. 
The arms host prominent dust lanes on their inner (concave) sides out to at
least $70\arcsec$. The two massive trailing CO spiral arms almost connect 
to the  $B$-band   arms and  can be followed from  $20\arcsec$ down  to 
$5\arcsec$. 
\label{fig2}}
\end{figure}

\clearpage
\begin{figure}
\centerline{
\psfig{figure=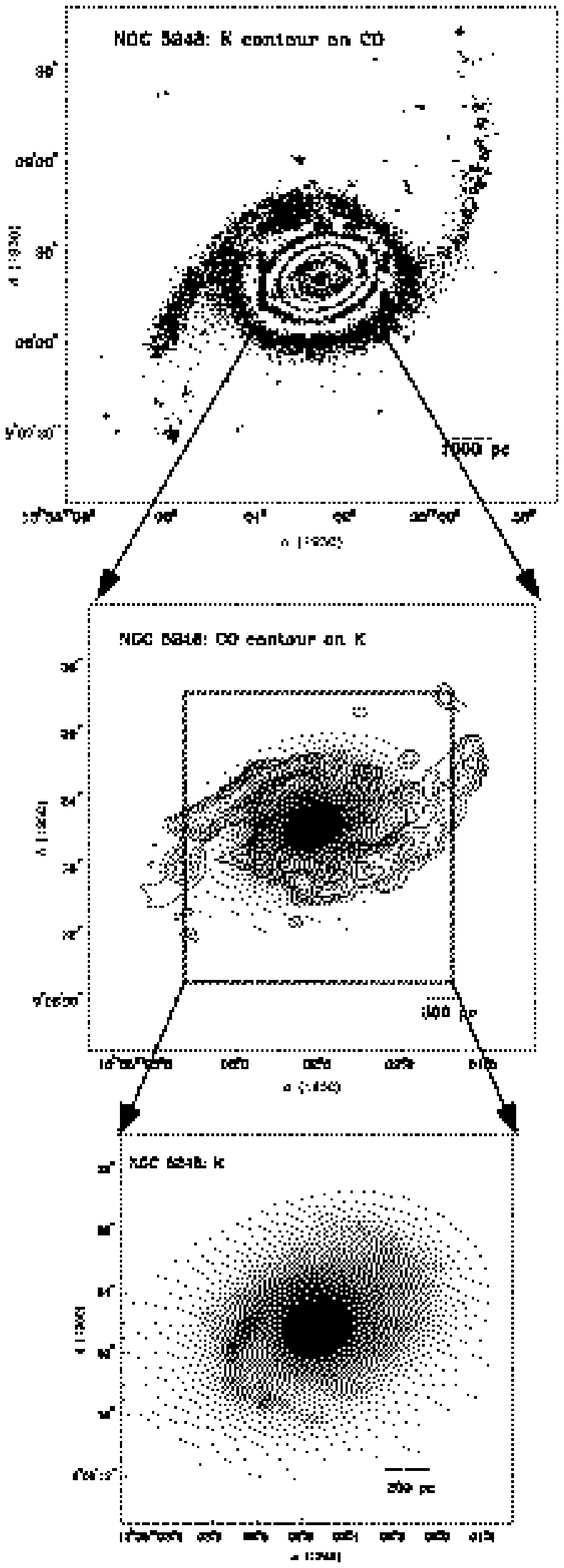,height=7.0in} 
}
\vspace{3pt}
\caption{
\bf
The  near-infrared morphology:
\rm 
\bf 
Top: 
\rm  
The CO (1--0) intensity map (gray-scale) of the central $40\arcsec$ 
is overlaid on a $K_{\rm s}$  image  (contours)  of the 
central $2\farcm 7$. 
The NIR image shows a weakly oval feature within the central $22\arcsec$  
(1.6~kpc) radius and prominent  $K$-band spirals which can be easily 
traced from  $85\arcsec$ down to  $\sim$~$26\arcsec$ where they cross the 
bar major axis. The  region  inside the box is enlarged below. 
\bf
Middle: 
\rm 
The CO (1--0) intensity (contours) is overlaid on the  $K_{\rm s}$ (gray-scale) 
image of  the central $40\arcsec$. 
\bf 
Bottom:  
\rm
The $K_{\rm s}$  (gray-scale) image  in the central $25\arcsec$ is shown. 
After the outer $K$-band spirals cross the bar major axis, only very 
faint  $K$-band emission is present between  $26\arcsec$ and $9\arcsec$. 
At $\sim$~$8\arcsec$,  two bright  inner $K$-band spirals  connect 
to the  CO arms, cross the starburst  ring while delineating the SSCs 
(see Figs. 6, 7, and 8), and   continue down to  $\sim$~$3\arcsec$.
\label{fig3}} 
\end{figure}

\clearpage
\begin{center}
\vspace{-3mm}
\includegraphics[height=3.75 in]{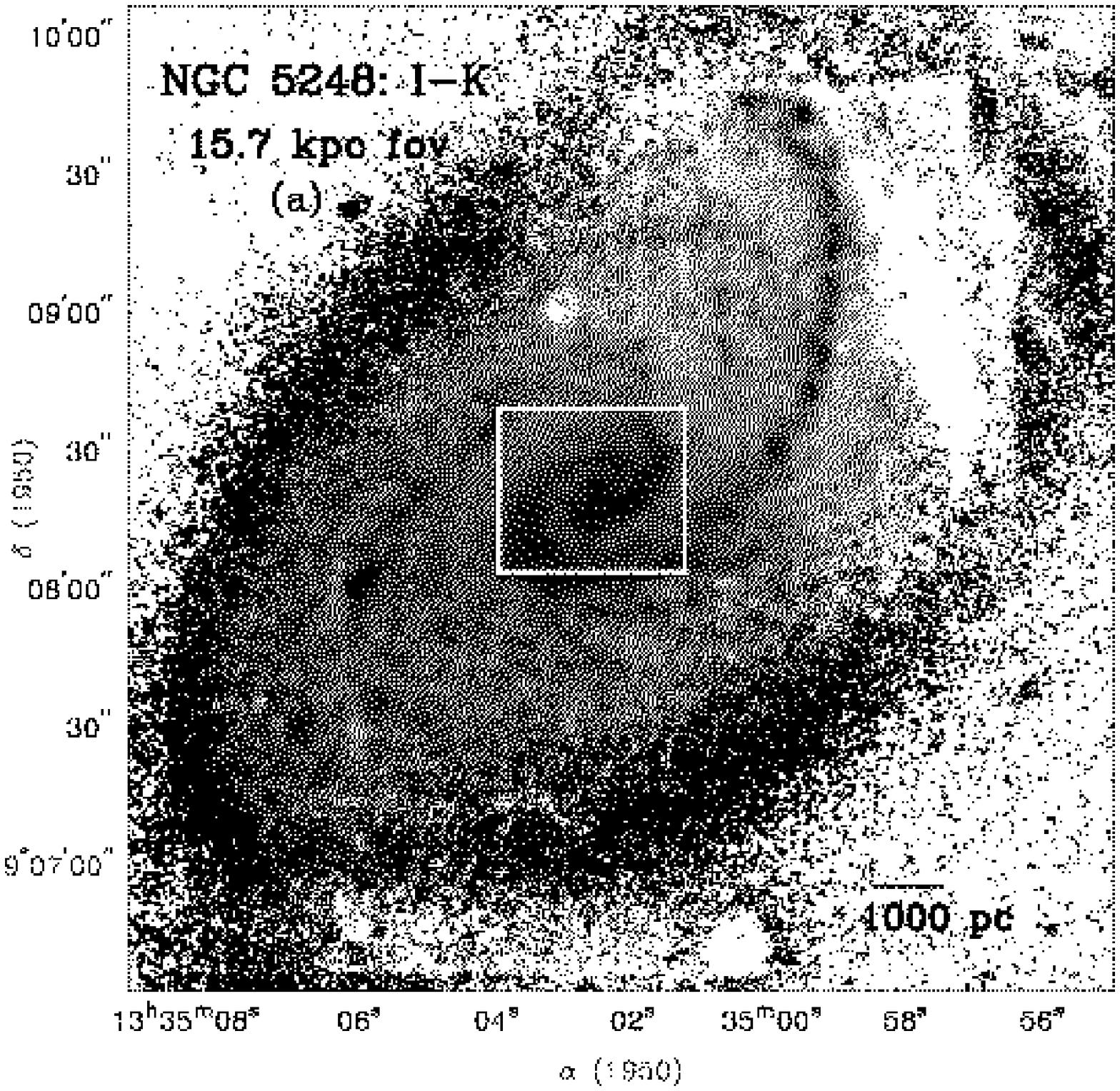}
\end{center}
\vspace{-8mm}
\begin{center}
\includegraphics[width =3.65 in]{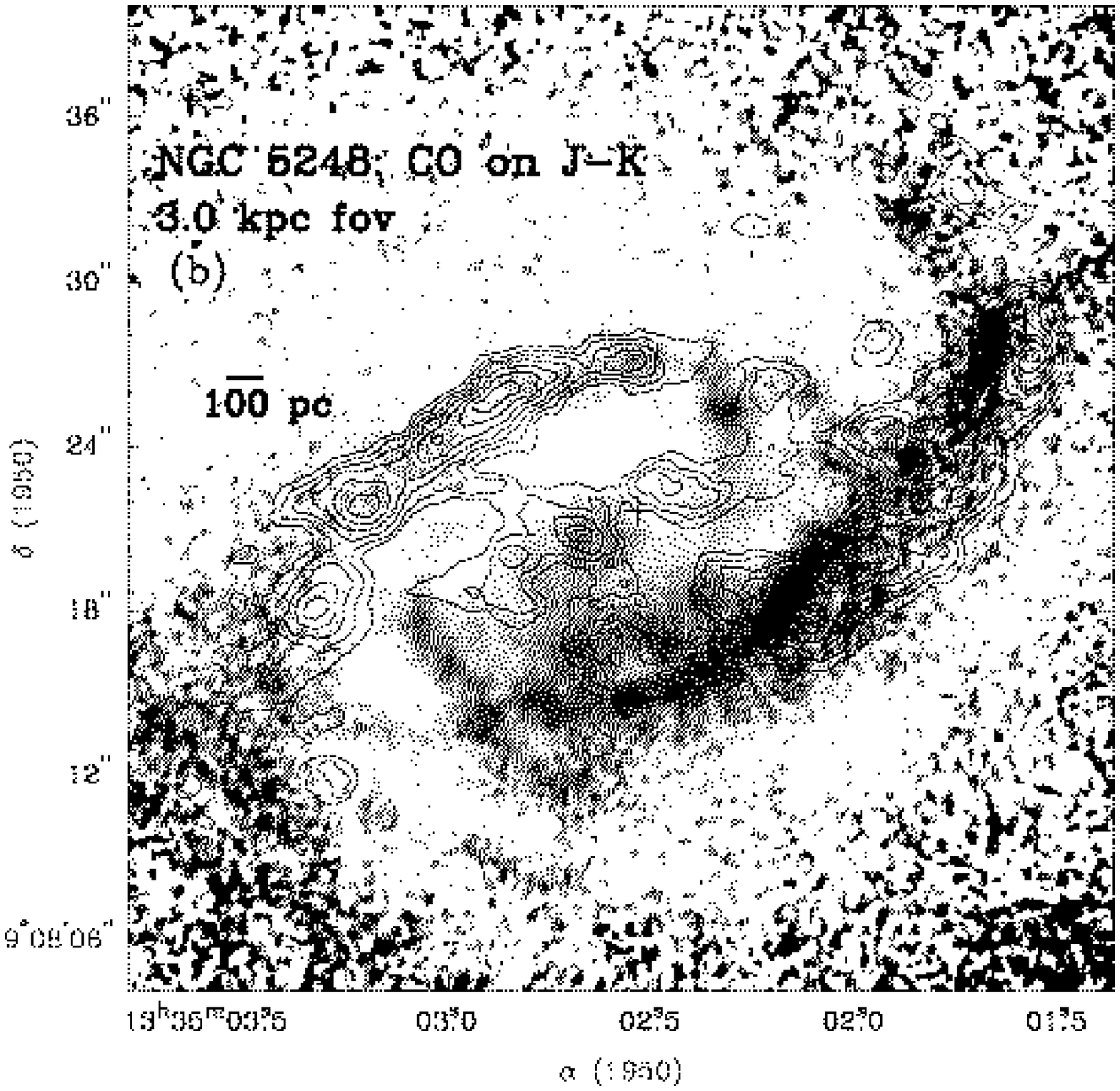}
\end{center}
\vspace{-10 mm}
\begin{description}
\item[]
\noindent
\rm
Fig. 4.--- \bf The grand-design dust spiral:  
\rm
\bf
(a)
\rm 
The $2\farcs 4$ resolution $I-K$ image shows a grand-design dust spiral
which is continuous from  at least $70\arcsec$ (5.2 kpc)  down to 
$5\arcsec$ (375 pc). The region inside the white box is shown in Fig 4b. 
\bf
(b)
\rm  
The CO (1--0)  intensity map  (contours) is overlaid  on a  
$1\farcs 5$ resolution   $J-K$ color map  (gray-scale) of the 
central 3.0 kpc. The  dust spiral follows the CO spiral arms between 
$20\arcsec$ and  $5\arcsec$.
\end{description}
\setcounter{figure}{4}

\begin{figure}
\plotone{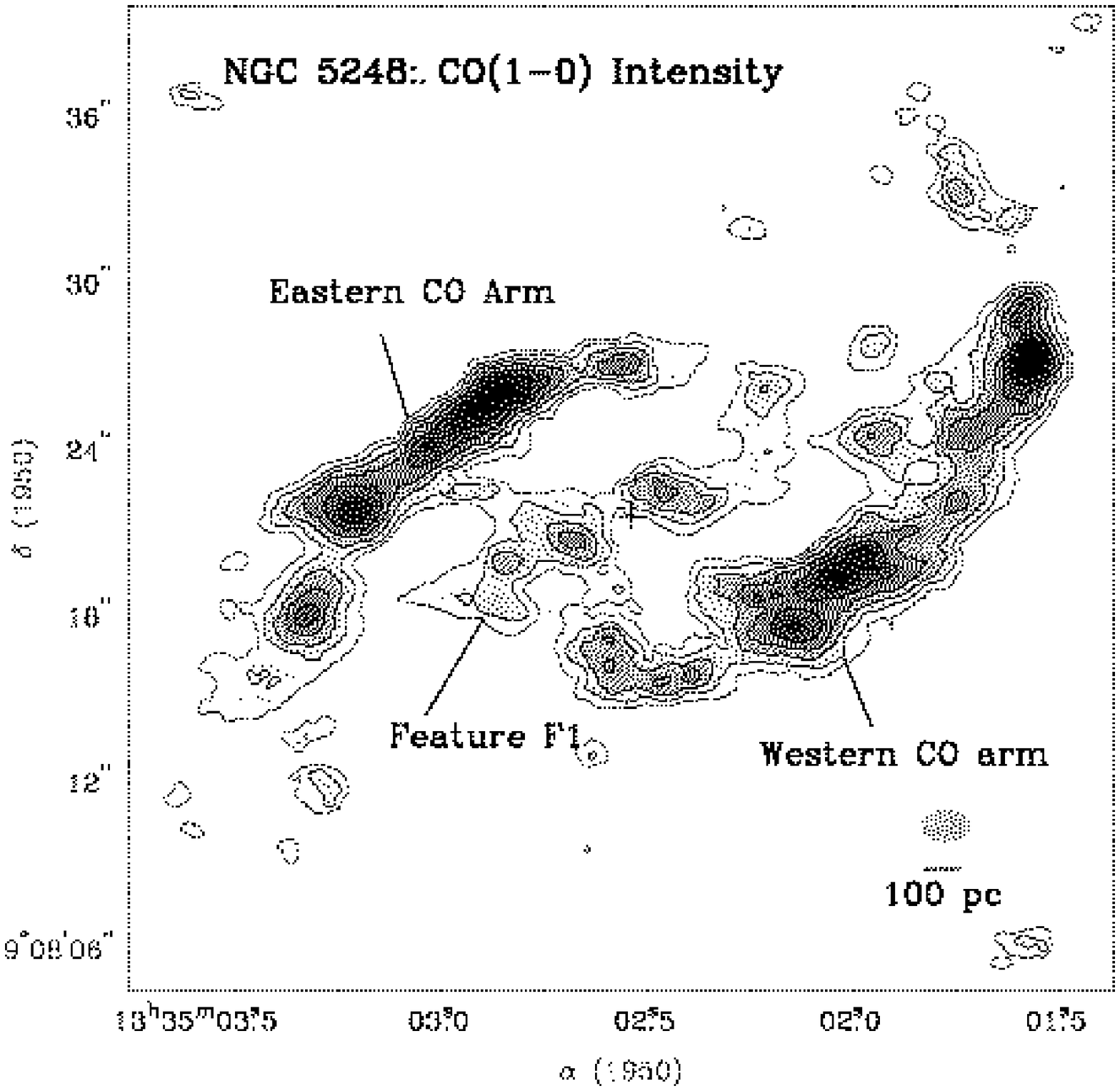}
\caption{
\bf
The molecular gas distribution in the inner kpc:
\rm
The CO (1--0) intensity map  of the central $40\arcsec$  is shown. 
The $1 \farcs 9 \times 1 \farcs 4$ ($140 \times 100$~pc) 
synthesized beam is marked. Contour levels are 
20, 35, 45, 55, 65, 75, 85, 95, and 100  \% of the peak flux density 
810 Jy beam$^{-1}$ km~s$^{-1}$, and show regions of bright emission.
The cross marks the peak of the {\it HST} $I$-band
(F814W) image. Two massive trailing  CO spiral arms  are seen  
from a radius of $20\arcsec$  to $5\arcsec$. 
The arms  are resolved into clumps with sizes ($2\arcsec$ to $4\arcsec$, 
or 150 to 300~pc) and masses  (several times $ 10^6$ to 10$^7$ M$_{\sun}$)
comparable to giant molecular associations. Further in resides 
a double-peaked CO feature surrounded by  fainter emission.  
\label{fig5}}
\end{figure}

\begin{figure}
\caption{
\bf
The distribution of CO and  H{\sc ii}  regions: 
\rm
This figure is included as a jpeg file. 
The CO (1--0)  intensity map  (contours) is overlaid  on the {\it HST} 
WFPC2 continuum-subtracted H$\alpha$+[N{\sc ii}]  image (gray-scale; 
courtesy of Maoz et al. 2001).
The two CO spiral arms  connect to  the circumnuclear  
starburst ring which has a  $5\arcsec$ radius and hosts 
H{\sc ii}  regions and SSCs. 
The H{\sc ii}  regions with   identification numbers 
3, 4, and 16  (labeled in red) have the largest extinction, 
and show no optically-visible  stellar continuum sources  although  
they have large Pa$\alpha$ equivalent widths indicative of a young  
embedded stellar population  a few  Myr  old. Interestingly, 
these complexes  are situated in  the CO arms and lie near 
bright CO peaks.  Conversely, the complexes 8, 14, 21, 12, 6, 17, 19, 7,
18, 20  (labeled in blue)  have the lowest extinction, typical ages of
10--40~Myr, and are associated  with lower levels of CO emission. 
Inside  the   circumnuclear   H$\alpha$  ring  resides a  second 
H$\alpha$ ring of radius $1 \farcs 25 $, which 
is crossed by the two CO peaks of feature F1. 
Northwest and southeast of the CO peaks, fainter CO emission 
extends towards the  $K$-band spirals in the  circumnuclear ring.
\label{fig6}}
\end{figure}

\begin{figure}
\caption{
\bf
The distribution of CO and  UV-bright SSCs: 
\rm 
This figure is included as a jpeg file. 
The CO (1--0)  intensity map  (contours) is overlaid  on the 
{\it HST}  F336W image (gray-scale).
Peaks in CO and UV emission are generally anticorrelated.
The stellar clusters labeled  SC11, SC12, SC13, and SC14 
lie along the narrow southern $K$-band spiral  (Fig. 8) which connects 
to the western CO arm. The clusters  SC21, SC22, and SC23 are more 
closely associated  with the northern $K$-band spiral (Fig. 8) which 
connects to the eastern CO arm. 
\label{fig7}}
\end{figure}

\begin{figure}
\caption{
\bf
The inner $K$-band spirals delineating the SSCs: 
\rm
This figure is included as a jpeg file. 
The  $K_{\rm s}$ image  depicting the same region  as Fig. 7 
is shown. The positions of the  UV-bright SSCs in the {\it HST}  UV (F336W) 
image (Fig 7) are marked. 
Two $K$-band spirals extend from $3\arcsec$  to $8\arcsec$.
The southern one is particularly clear and narrow,  suggesting
a young dynamical age. There is a striking spatial coincidence  between 
emission peaks along the $K$-band spirals and the UV-bright SSCs.
\label{fig8}}
\end{figure}

\begin{figure}
\plotone{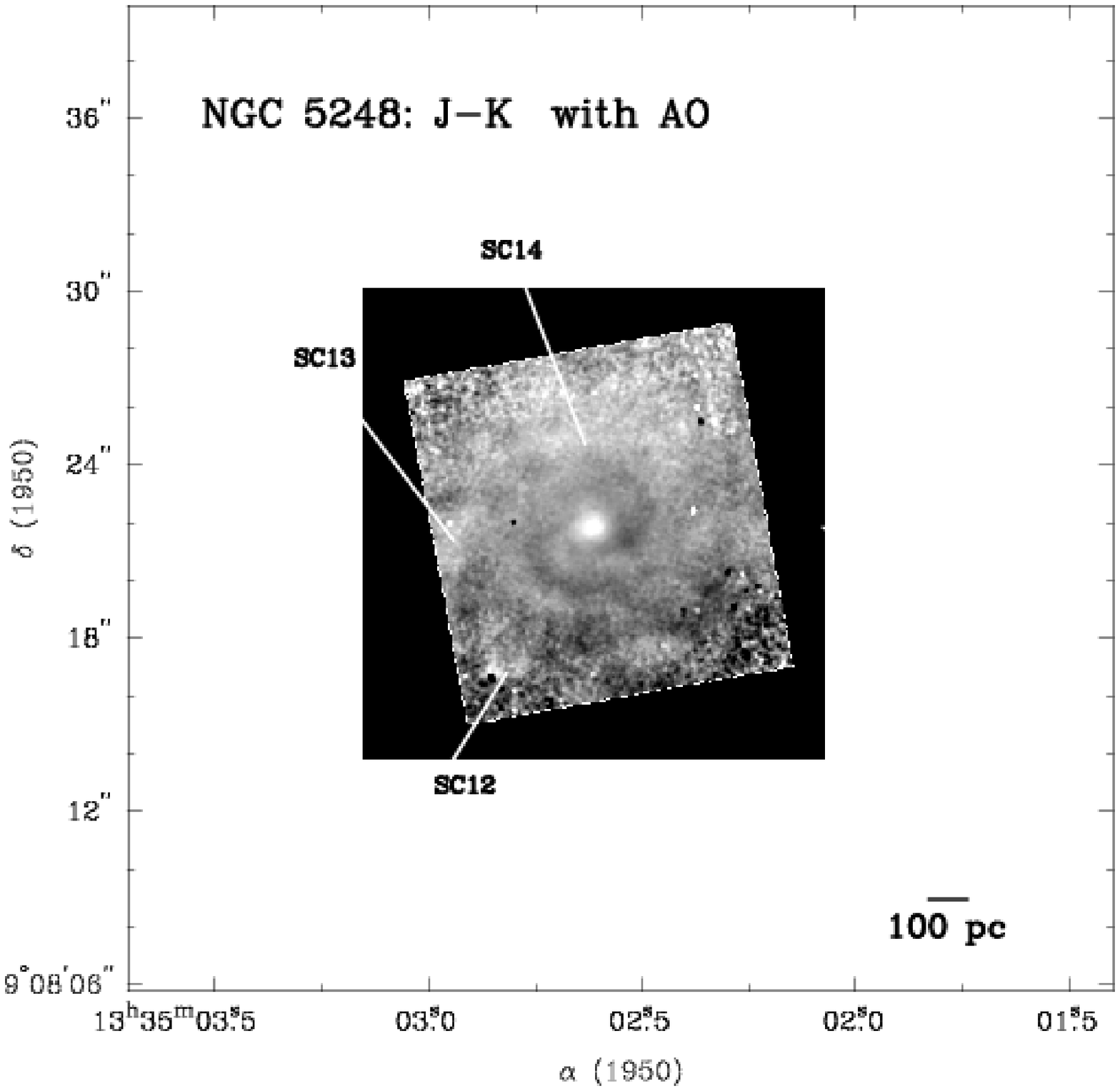}
\caption{
\bf
The nuclear grand-design dust spiral:  
\rm 
The CFHT adaptive optics $J-K$ color index image (Laine et al. 1999a) 
of the core region of NGC 5248  shows the nuclear  
grand-design dust spiral in darker shades.
The positions of three UV-bright SSCs lying along the narrow 
western $K$-band arm, shown in Figs. 7 and 8,  are marked.
Notice that the northern dust spiral appears to connect 
to  this $K$-band arm around $3\arcsec$. 
\label{fig9}}
\end{figure}

\clearpage
\begin{center}
\includegraphics[height=3.8in]{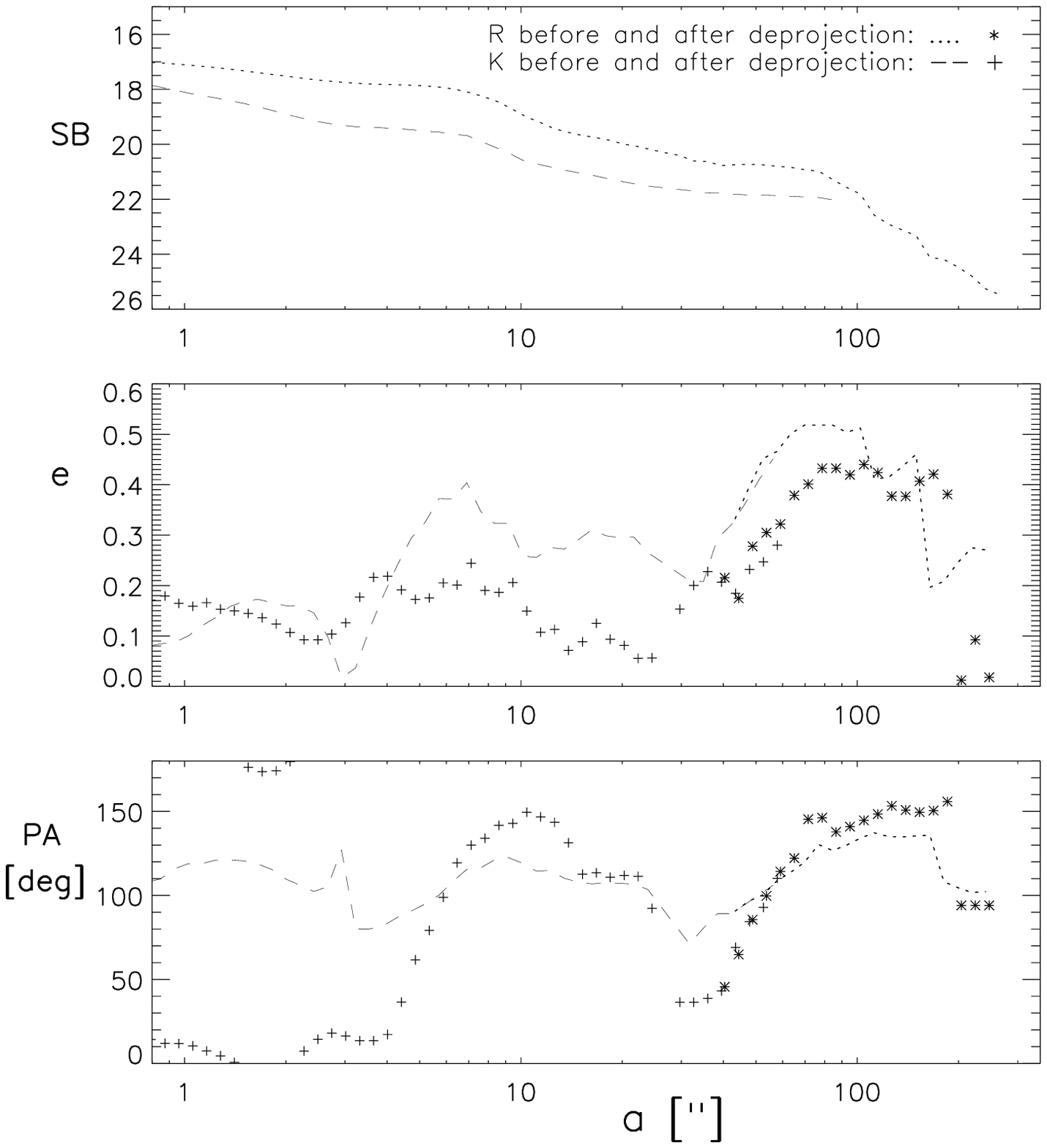}
\end{center}
\vspace{-3 mm}
\hspace{35 mm}
\includegraphics[height=3.93 in]{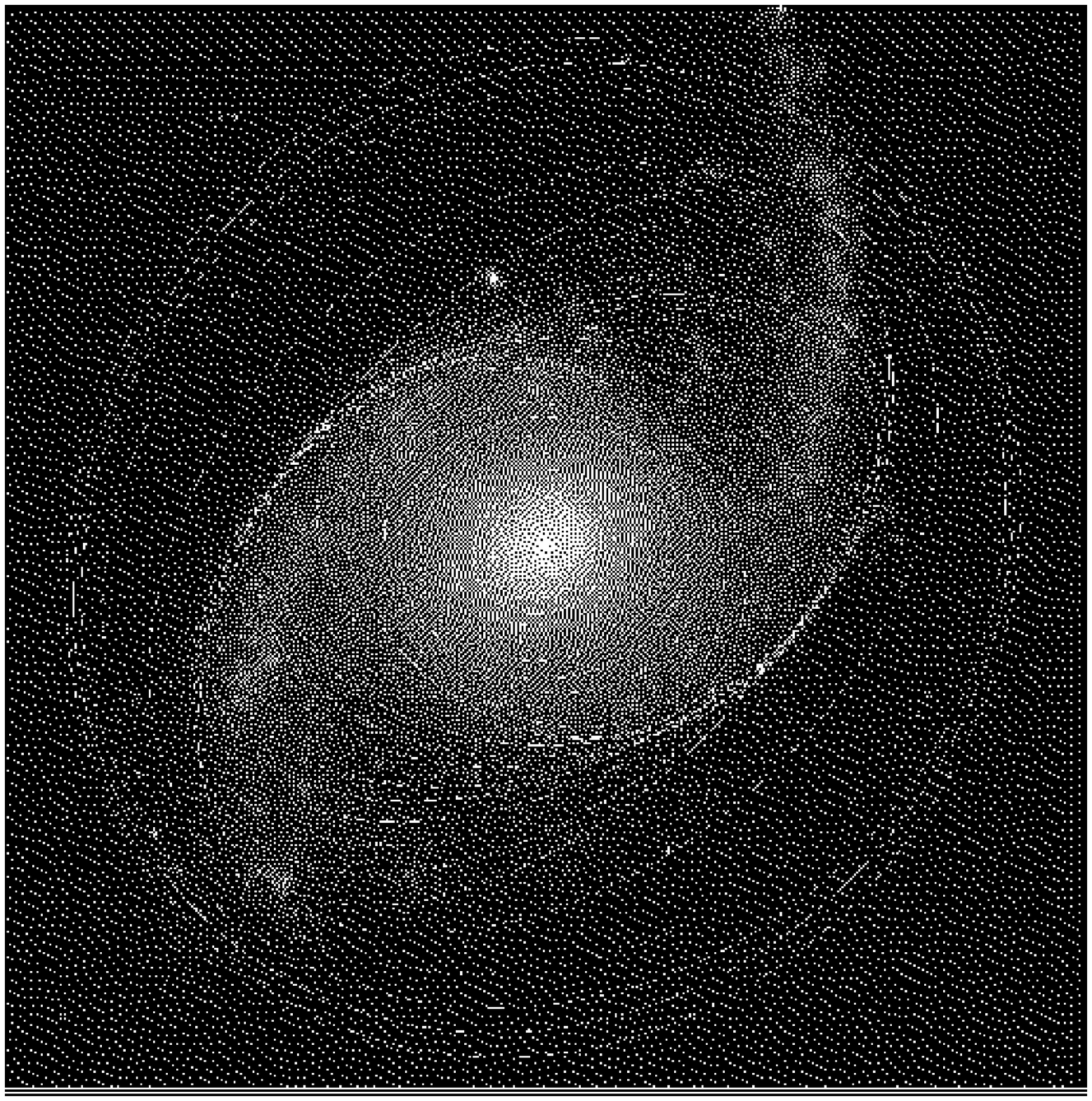}
\vspace{-4 mm}
\begin{description}
\item[]
\noindent
\rm
Fig. 10--- 
\bf
Isophotal analysis:
\rm 
\bf
(a) 
\rm 
Top: The radial profiles of surface brightness (SB), ellipticity ($e$),  
and  position angle (PA) of the  $K_{\rm s}$
 and $R$-band  light are shown. 
The dotted and dashed lines refer to the values before deprojection.
The symbols  refer to  deprojected values.
The $K_{\rm s}$  SB has been scaled by 1.32 so that
it can be plotted on the same scale as the  $R$-band profile.
In  deprojected images, we adopt the convention that PAs 
are measured anticlockwise from ``North''.
The gradual change in PA (isophotal twist) between 
$3\arcsec$ and $9\arcsec$ is caused by  $K$-band arms.
\bf
(b) 
\rm 
Bottom: The fitted isophotes  to the deprojected  $K_{\rm s}$  image 
provide a  guide to the  dominant periodic stellar 
orbits present and the location of dynamical resonances. 
\end{description}
\setcounter{figure}{10}

\clearpage
\begin{center}
\includegraphics[height=4.2in]{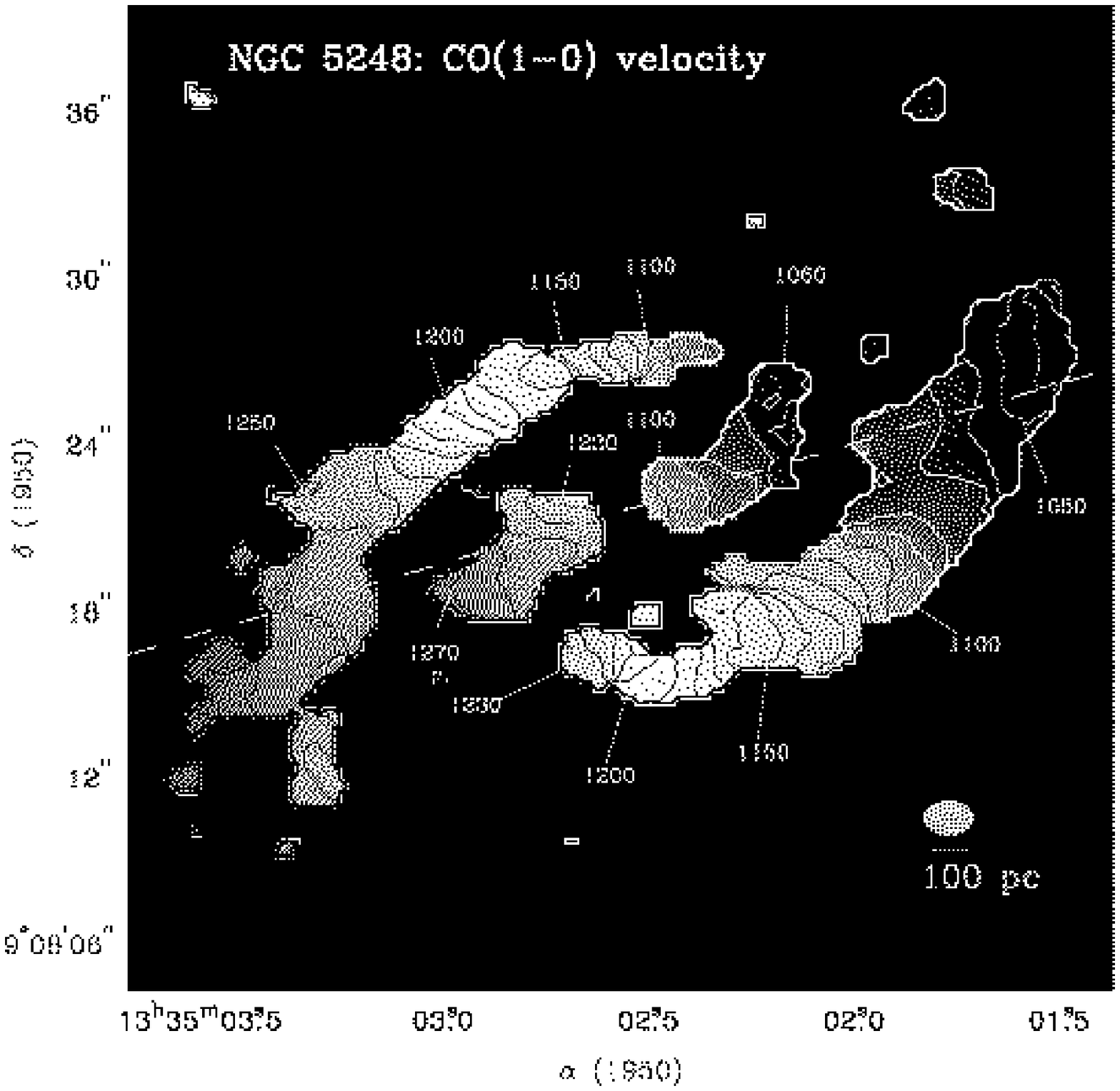}
\end{center}
\hspace{-4mm}
\includegraphics[height=3.1in]{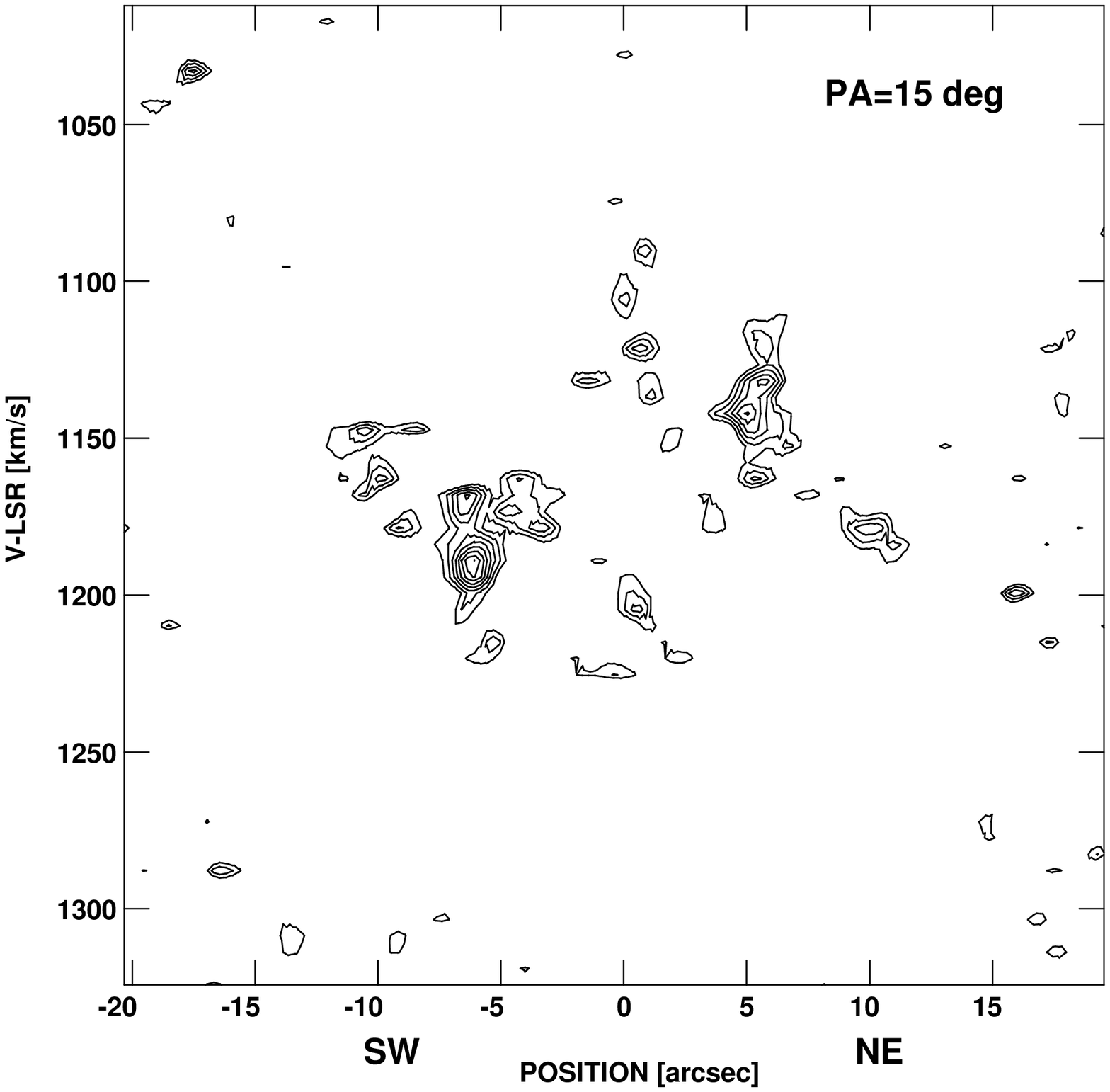}
\hspace{8mm}
\includegraphics[width=3.1in]{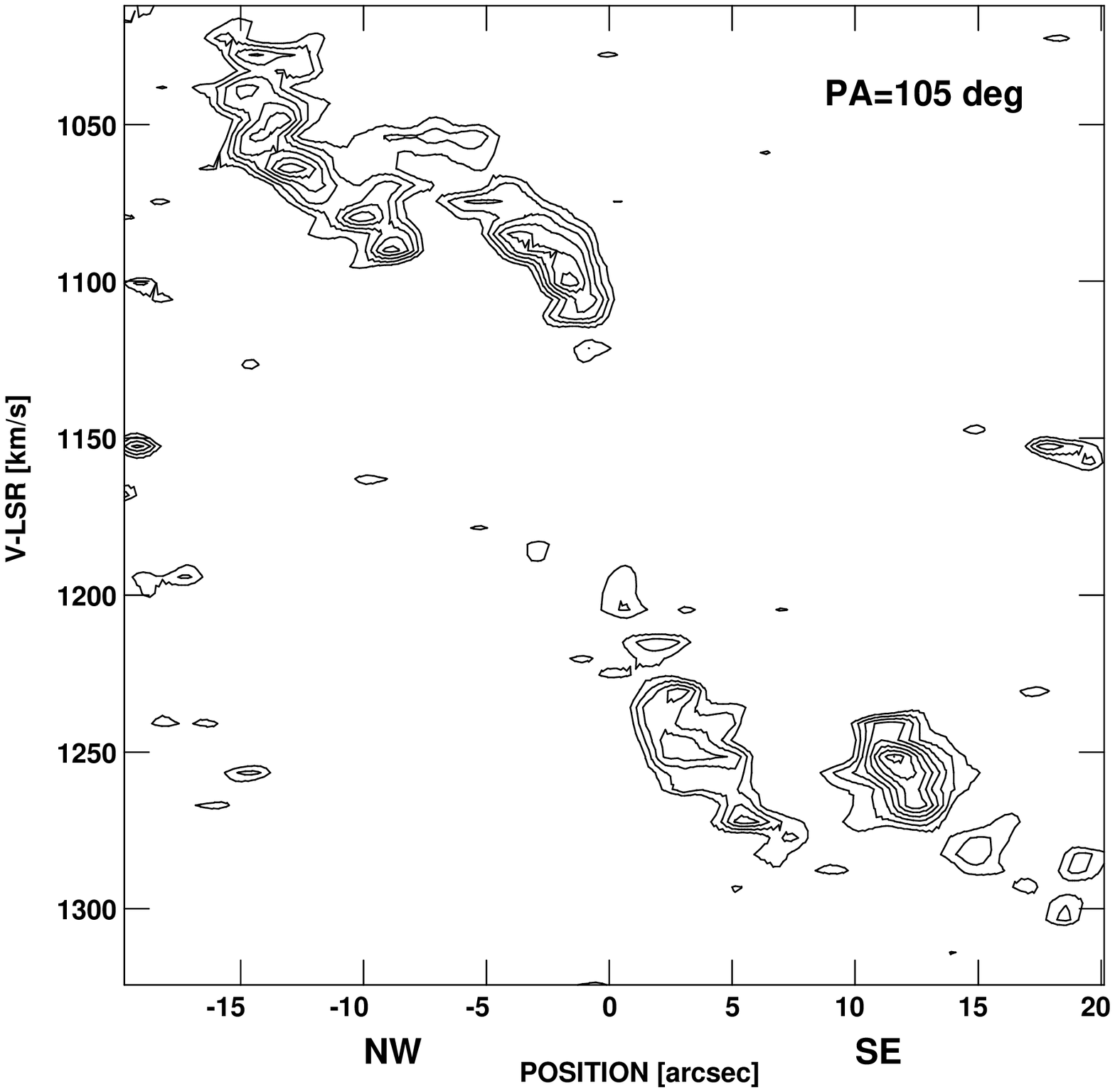}
\vspace{-8 mm}
\begin{description}
\item[]
\noindent
\rm
Fig. 11--- 
\bf
The molecular gas kinematics:
\rm 
\bf (a) \rm  Top: 
The CO (1--0) intensity-weighted velocity field 
of the central $40\arcsec$  is shown. 
The $1 \farcs 9 \times 1 \farcs 4$ 
synthesized beam is marked. Contour levels range from 1000 to  
1300~km~s$^{-1}$ and  are plotted at intervals of  10~km~s$^{-1}$. 
\bf (b) \rm  Bottom left: 
The CO  p--v  cuts along the minor axis ($15^\circ$).
\bf (c) \rm Bottom right:  
The CO p--v cut  along the  line of nodes ($105^\circ$). 
\end{description}
\setcounter{figure}{11}

\begin{figure}
\plotone{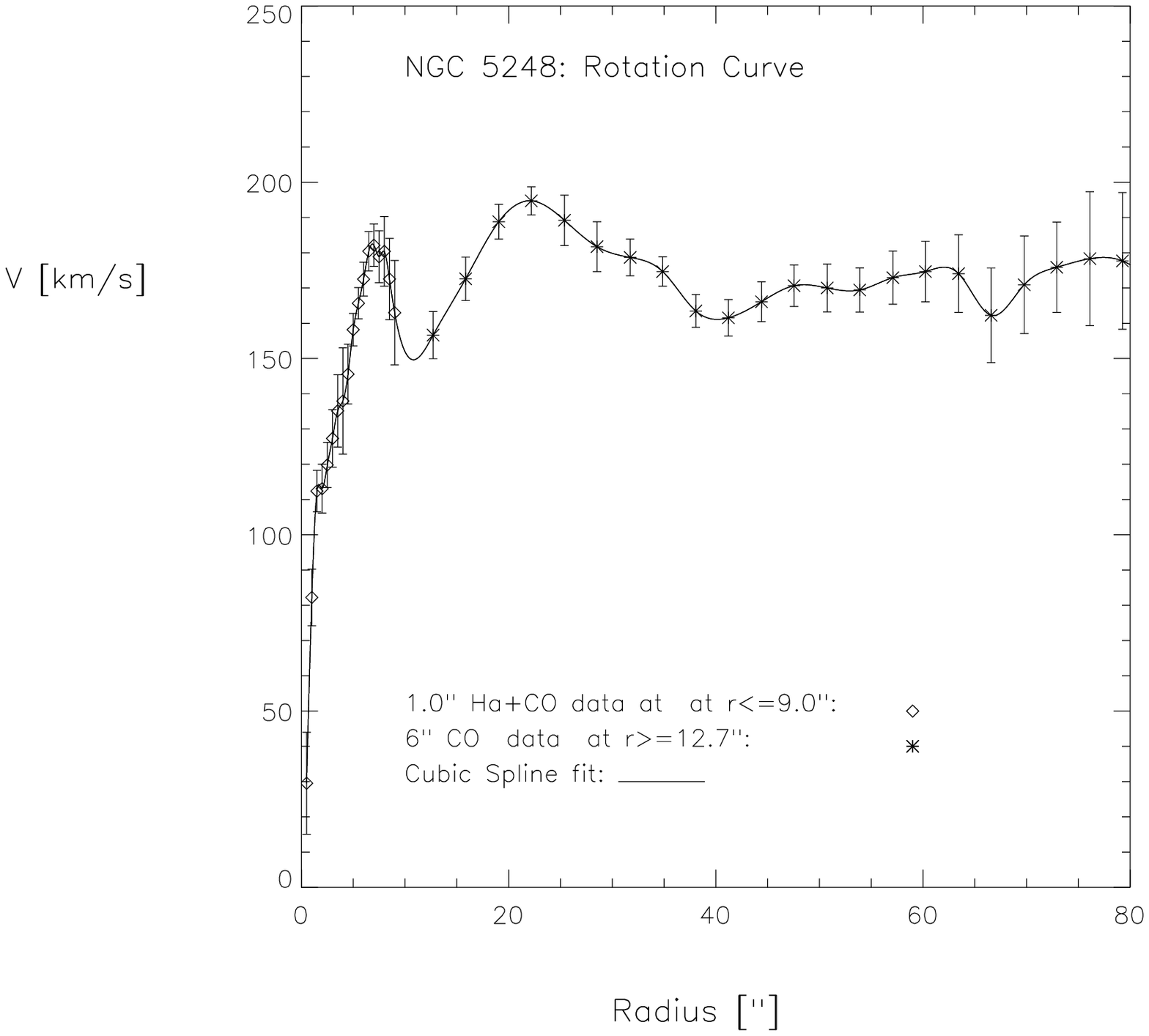}
\caption{
\bf The rotation curve: \rm
Shown is the   rotation curve  
from the center  to $80\arcsec$, already corrected for an assumed 
inclination of  $40^\circ$.  It  is based on the high resolution  
Fabry--Perot H$\alpha$ and OVRO CO (1--0) data  out  to $10\arcsec$, 
and BIMA data of lower resolution from $13\arcsec$ to $80\arcsec$.
We interpolated between data points with a  cubic  spline fit.  
The rotation curve is likely contaminated by non-circular motions 
between  $8\arcsec$ and  $11\arcsec$ and  $12\arcsec$ and $18\arcsec$  
(see text;$\S$ 5).
\label{fig12}}
\end{figure}

\clearpage
\vspace{-8 mm}
\begin{center}
\end{center}
\vspace{-2 mm}
\begin{description}
\item[]
\noindent
\rm
Fig. 13--- 
\bf 
Comparison of the CO and $R$-band spirals with 
hydrodynamical models of bar-driven gaseous spiral density waves in 
NGC 5248. \rm 
This figure is included as a jpeg file. 
\bf Left: Models --- \rm 
The steady-state gas response  in models  of 
bar-driven gaseous SDW  is shown 
\bf  (a) \rm from 
the central regions out to the end of the bar, 
and \bf (c)  \rm  only in the inner few kpc, deep inside the OILR 
of the bar.  Points P1' to P4'  in the models  correspond to points 
P1 to P4 in the data. The  orientation of the large-scale stellar bar 
is shown by a dotted line. Strong bar shocks around the OILR 
excite a trailing  non-linear  high amplitude gaseous SDW which weakens 
rapidly as it travels inside the OILR. Under the right conditions,  it 
can excite a  linear  low amplitude  gaseous SDW  inside a transition 
radius $R_{\rm t}$.  This linear   SDW  can wind through an arbitrary 
angle, because its shape is not dictated by the dominant stellar  orbits. 
\bf Right: The  data --- \rm 
The  deprojected CO map  (white contours) of the central $40\arcsec$
 (3.0 kpc) is overlaid on a  large-scale $R$-band image. 
Regions of size \bf (b) \rm   $8 \arcmin $  (36 kpc) and 
\bf (d) \rm  $2 \arcmin$ (9  kpc)  are shown.  The  two massive  trailing 
CO spiral arms lie deep inside the OILR, cover more than 180 $^\circ$ in 
azimuth, and  appear to correspond to the high amplitude non-linear \
gaseous SDW. The  double-peaked CO feature  F1 and the nuclear dust spiral 
which  lie  inside $4\arcsec$  appear to be associated with 
the linear gaseous SDW inside the transition radius  $R_{\rm t}$. 
\end{description}
\setcounter{figure}{13}

\clearpage
\vspace{-8 mm}
\begin{center}
\includegraphics[height=6.3in]{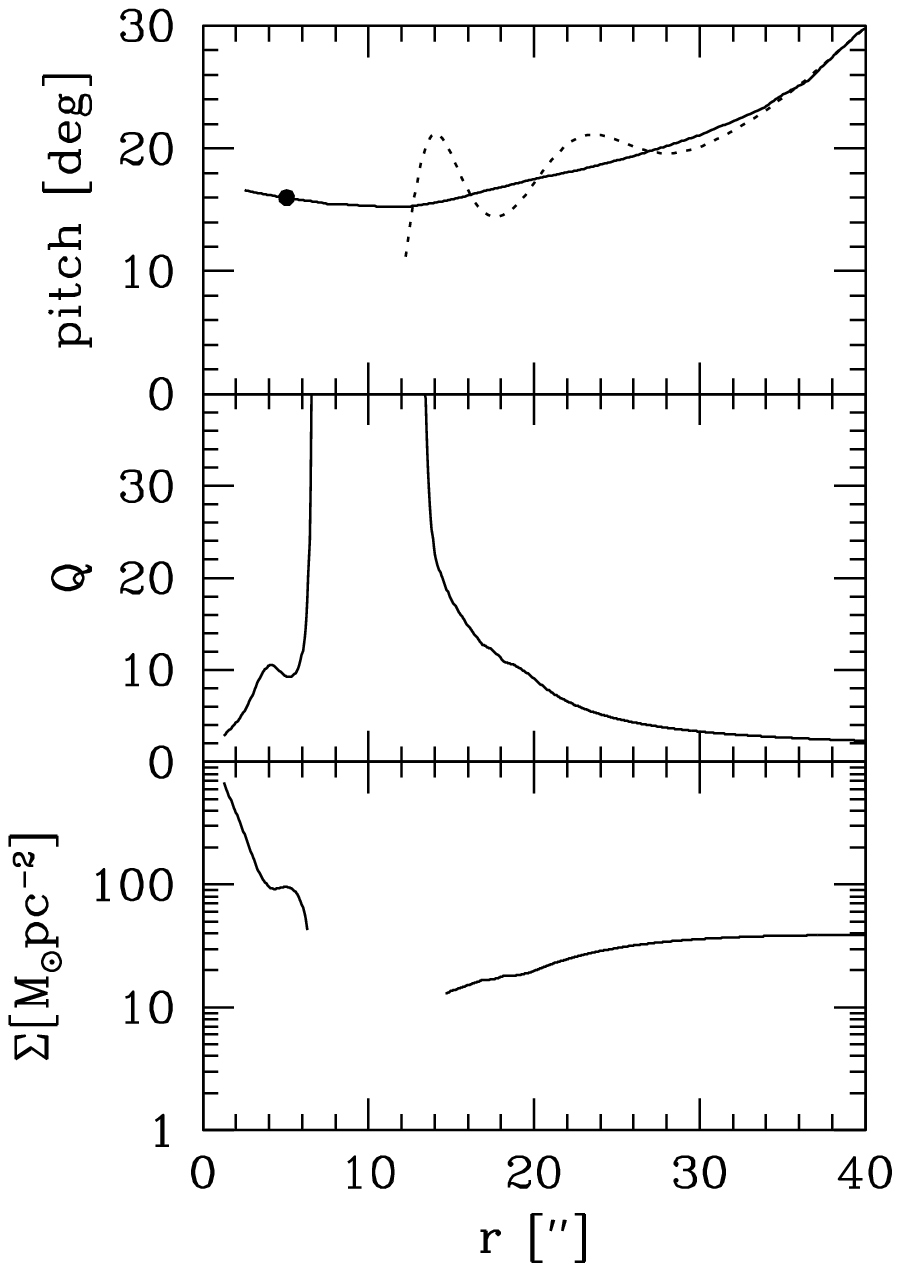}
\end{center}
\vspace{-8 mm}
\begin{description}
\item[]
\noindent
\rm
Fig. 14--- 
\bf
Comparison of observed and model pitch angles: 
\rm 
\bf (a) \rm  Top: 
The observed pitch angle for the nuclear dusty spiral (filled circle) and the
CO spirals (dotted line), as well as, the best eyeball-fit to these data (solid
line). For the eyeball-fit, we computed the model including self-gravity term
shown in the other two  panels based on the rotation curve of NGC~5248, a sound
speed  $v_{\rm s} = 16$~km~s$^{-1}$ and a pattern speed $\Omega_{\rm s}$ of
25~km~s$^{-1}$ kpc$^{-1}$. 
\bf (b) \rm Middle: 
Effective Toomre parameter $Q$ (see the text) along the spiral arms.
\bf (c) \rm Bottom: 
Inferred gas surface density profile along the spiral arms. A transition
radius, $R_{\rm t}$, separates the outer non-linear  gas response from the
linear response at smaller radii.  
\end{description}
\setcounter{figure}{14}

\clearpage
\vspace{-8 mm}
\begin{center}
\includegraphics[height=6.1in]{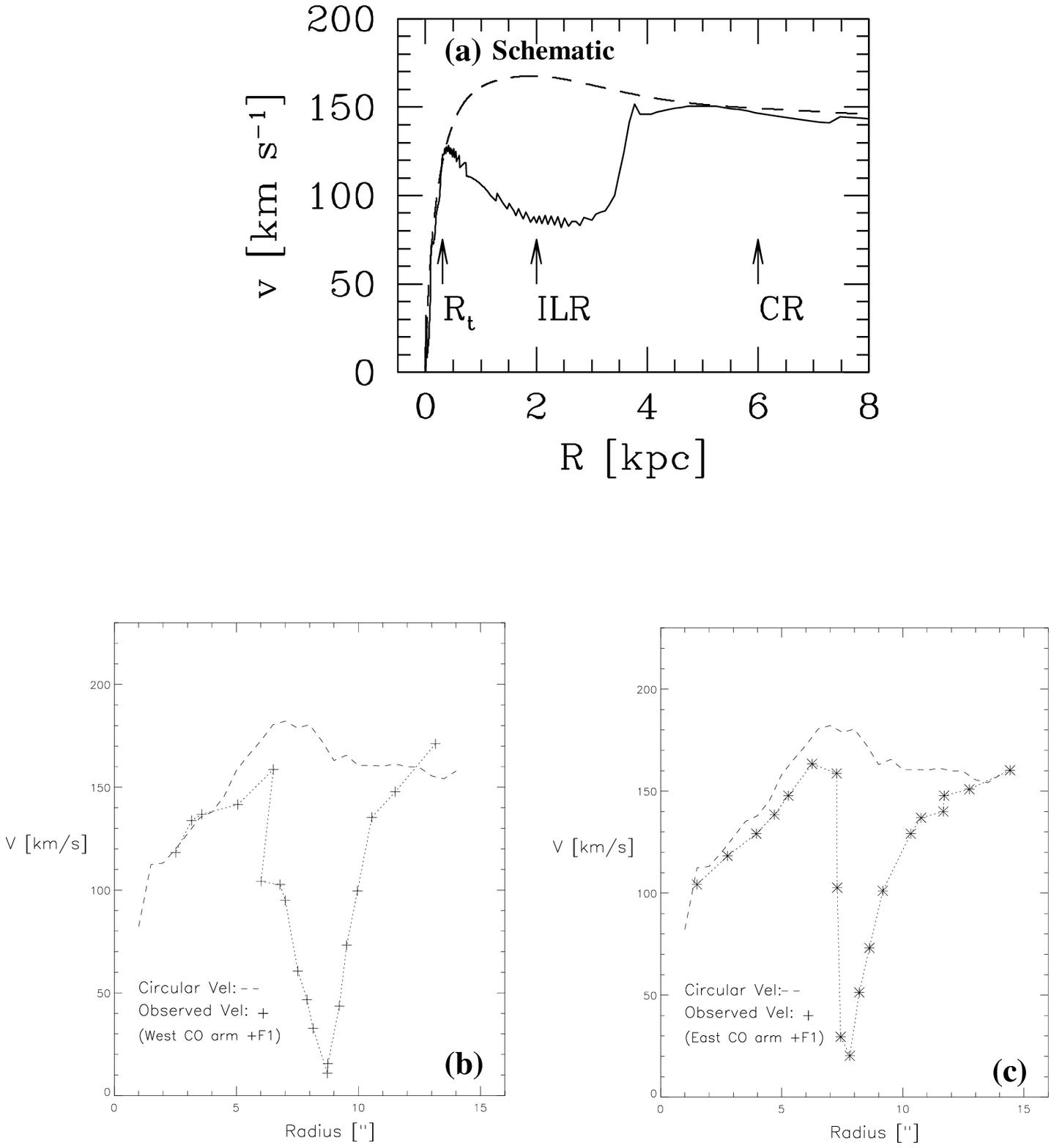}
\end{center}
\vspace{-8.0 mm}
\begin{description}
\item[]
\noindent
\rm
Fig. 15--- 
\bf
Comparison of observed and model kinematics: 
\rm
\bf
(a) 
\rm
The gas kinematics predicted by  models of the gaseous SDW are shown.
Close to the OILR, where shocks from 
the bar perturb most, large deviations from purely circular 
motions are expected. Inside the transition radius $R_{\rm t}$ the  
azimuthal speed   is very close to the circular speed 
because  the $x_{\rm  2}$   orbits are almost circular. 
A sharp change in behavior is expected at  $R_{\rm t}$. 
\bf
(b) 
\rm
The observed difference between  the  circular speed curve and 
the CO velocity is shown. The velocity is measured 
along the western trailing CO spiral  arm, and the  
southern part of the double-peaked CO feature F1.
The largest deviations from circular motions occur 
between  $7\arcsec$ and $10\arcsec$  and a sharp change is 
seen at  a radius of about $6\arcsec$  
where the observed and circular speeds  converge.
\bf
(c) 
\rm
Same as in (b), but for the eastern trailing CO spiral  arm, 
and the northern part of the CO feature F1.
\end{description}
\setcounter{figure}{15}

\end {document}